\documentclass[twocolumn,amsmath,amssymb]{revtex4}

\usepackage{graphicx}
\usepackage{dcolumn}
\usepackage{bm}

\newcommand{\re}{\mathop{\mathrm{Re}}}
\newcommand{\im}{\mathop{\mathrm{Im}}}

\begin{document}

\title{Topics on High-Energy Elastic Hadron Scattering }

\author{M. J. Menon}

\affiliation{
Instituto de F\'{\i}sica ``Gleb Wataghin''\\
Universidade Estadual de Campinas, UNICAMP\\
13083-970, Campinas, SP, Brazil
}

\date{\today}%

\begin{abstract}
We review the main results we have obtained in the area of
high-energy elastic hadron scattering and presented in this
series of \textit {Workshops on Hadron Interactions}.
After an introduction to some basic experimental and theoretical concepts,
we survey the results reached by means of four 
approaches: analytic models, model-independent analyses,
eikonal models and nonperturbative QCD. Some of the ongoing researches
and future perspectives are also outlined.
\end{abstract}

\maketitle

\centerline{\bf Contents}

\vspace{0.3cm}

\leftline{\bf I. Introduction}

\vspace{0.3cm}

\leftline{\bf II. Basic Concepts}

II.A Physical Quantities

II.B Principles, Theorems and High-energy Bounds

II.C Basic Pictures

\ \ \ II.C.1 Optical/Geometrical Picture

\ \ \ II.C.2 Exchange Picture
                      
\vspace{0.3cm}

\leftline{\bf III Analytic Approach}

III.A Derivative Dispersion Relations

III.B Basic Models

\ \ \ III.B.1 Ensembles

\ \ \ III.B.2 Analytic Models

\ \ \ III.B.3 Fits and Results

III.C Non-degenerate Meson Trajectories

\ \ \ III.C.1 Extrema Bounds for the Pomeron Intercept

\ \ \ III.C.2 IDR, DDR and the Subtraction Constant

\vspace{0.3cm}

\leftline{\bf IV Model Independent Analyses }

IV.A Differential Cross Sections

\ \ \ IV.A.1 Unconstrained Fits and the Eikonal

\ \ \ IV.A.2 Constrained Fits  and Energy Dependence

IV.B Total Cross Sections and Slopes

\vspace{0.3cm}

\leftline{\bf V Eikonal Models}

V.A Geometrical Model - Inelastic Channel

V.B QCD-inspired Models

\ \ \ V.B.1 Basic Formalism

\ \ \ V.B.2 Dynamical Gluon Mass

\ \ \ V.B.3 Momentum Scale

\vspace{0.3cm}

\leftline{\bf VI Non-perturbative QCD}

VI.A Stochastic Vacuum Model

VI.B Elementary Amplitudes

\ \ \ VI.B.1 Correlators from Lattice QCD

\ \ \ VI.B.2 Correlators from Instanton Approach

\vspace{0.3cm}

\leftline{\bf VII Perspectives and Outlook}

VII.A Experiments

VII.B Theory

\vspace{0.3cm}

\leftline{\bf VIII Summary and Final Remarks}

\vspace{0.3cm}

\leftline{\bf References}

\vspace{0.3cm}

\section{Introduction}

\begin{quote}
 ``QCD nowadays has a split personality. It embodies hard
and soft physics, both being hard subjects and
the softer the
harder.''
\end{quote}
\centerline{Yuri Dokshitzer (2001) \cite{dokshitzer}}

\vspace{0.2cm}

Despite the great success of QCD as the field theory of hadronic
interactions, there still remains some open questions and
one of them is related to the hadron-hadron scattering at 
\textit{high-energies} and
\textit{small momentum transfer} (soft diffraction).

The region of high energies is characterized by scattering
of particles with center of mass energy 
$\sqrt s > 10$ GeV $\sim 10$ $m_p$ (the proton mass).
From the experimental point of view, diffractive processes
are associated with
a slow increase of the total cross sections,
the diffraction pattern in the differential cross section,
and rapidity gaps in the plots of pseudo rapidity {\it versus} 
azimuthal angle. In the theoretical context, diffraction means
that
the initial and final states in the scattering process
have the same quantum numbers and, therefore, the
exchanged ``object" has the vacuum quantum numbers
(Pomeron). The soft diffractive processes are generally classified as double 
diffraction dissociation, single diffraction dissociation and 
elastic scattering. Introductory reviews on the area can be found in Refs. 
\cite{bc,matthiae,engel,pred,ddln,menon02}.

High-energy elastic hadron scattering is the simplest soft
diffractive process and, at the same time a topical problem in 
high-energy physics. Being associated with long distance
phenomena perturbative QCD can not
be applied. On the other hand,
the standard non-perturbative approach starts with the ground state 
(vacuum), proceeds with bound states (mesons, barions) and eventually
reaches the scattering states. However, it is obvious that the vacuum  
is a non-trivial 
problem. 
Moreover, even assuming some vacuum concept, to treat  
only one gluon field 
it is necessary to take into account more than 30 invariants, and 
all that 
becomes
a typical problem of statistical physics, with specific technical 
approaches, 
such as Monte Carlo simulation (lattice QCD). Although bound
states may be described,
the point is that,
presently, we do not know how
to calculate elastic scattering amplitudes from a pure nonperturbative
QCD formalism. 

At this stage, phenomenology certainly plays an important role in the 
search for
connections between experimental data, model descriptions, and the
possible development of new calculational schemes in the underlying
theory (QCD). Here, however, we are faced with another kind of 
problem, namely, the wide variety of model descriptions, based on
different ideas and approaches, not always giving enough support
for the development of novel calculational schemes well founded on
QCD.

Based on the above facts, 
our main strategy in the investigation
of the elastic sector
is to search for \textit{ model independent information} 
that may
be extracted from the experimental data, through approaches that
have well established bases on the General Principles, theorems
and bounds from axiomatic quantum field theory
(the \textit{analytic approach}). Simultaneously,
we attempt to construct phenomenological models, in agreement
with the above Principles and connected,
in some way, with the underlying
dynamics of QCD.

In this review, it is presented some results we have obtained in the
area of elastic scattering in the last years, with focus on high-energy 
proton-proton 
($pp$) and antiproton-proton 
($\bar{p}p$) elastic scattering.
The manuscript is organized as follows. In Sec. II we recall some
basic experimental and theoretical concepts,
defining also our notation. In Secs. III, IV, V, and VI we present the
main
results we have obtained throughout the analytic approach, model
independent analyses, eikonal models and 
nonperturbative QCD, respectively. In Sec. VII we discuss
some perspectives in the area, from both experimental and theoretical
points of view.
A summary and some final remarks are the contents of Sec. VIII.  
 
\section{Basic concepts}

In this section we recall the physical quantities that characterize
the elastic scattering and shortly review some principles, high-energy
theorems, and the main formulas associated with two basic pictures,
usually referred as s-channel (geometrical/optical picture) and t-channel
(exchange picture) \cite{matthiae,engel,pred,ddln,menon02}.

\subsection{Physical Quantities}

In elastic scattering, the connection between experimental data and theory
is done by means of the \textit{invariant scattering amplitude}, 
expressed in terms of
two Mandelstam variables, generally the 
center-of-mass (c.m.) energy squared $s$ and the 
four-momentum
transfer squared $t = -q^2$: $F = F(s,t)$.
It is expected that spin effects decrease as the energy increases
(for some recent results see \cite{mp}), and
neglecting spin, the physical quantities that characterize the
elastic scattering process are the
differential cross section,

\begin{eqnarray}
\frac{d\sigma}{dt}(s,t) = \frac{\pi}{k^2}|F(s,t)|^2, 
\end{eqnarray}
where $k$ is the c.m. momentum, the elastic integrated cross section,

\begin{eqnarray}
\sigma_{el}(s) = \int_{-\infty}^{0} \frac{d\sigma}{dt}(s,t) dt, \nonumber
\end{eqnarray}
the total cross section (Optical Theorem),

\begin{eqnarray}
\sigma_{tot}(s) = \frac{4\pi}{k}\im F(s,0), 
\end{eqnarray}
the inelastic cross section

\begin{eqnarray}
\sigma_{inel}(s) = \sigma_{tot}(s) - \sigma_{el}(s), \nonumber
\end{eqnarray}
the $\rho$ parameter,

\begin{eqnarray}
\rho(s) = \frac{\re F(s,0)}{\im F(s,0)}, 
\end{eqnarray}
and the slope parameter,

\begin{eqnarray}
B(s) = \frac{d}{dt} \left[ \ln \frac{d\sigma}{dt}(s,t) \right]_{t=0}.
\end{eqnarray}

The corresponding experimental data have been analyzed and
compiled by the Particle Data Group and can be found in Ref.
\cite{pdg} and quoted references. In what follows we shall be
mainly interested in $pp$ and $\bar{p}p$ data in the regions:
$13.8$ GeV $\leq \sqrt s \leq 1.8$ TeV and
$0.01$ GeV$^{2} \leq q^{2} \leq 9.8$ GeV$^{2}$.
In a particular analysis we shall also use the
$pp$ data at $\sqrt s = 27.5$ GeV, in the
region $5.5$ GeV$^{2} \leq q^{2} \leq 14.2$ GeV$^{2}$.
Some treatment of cosmic-ray information on $pp$ total
cross sections at $\sqrt s =$ 6 - 40 TeV is also
presented.

In Figure 1 it is displayed the
experimental information available on $pp$ and $\bar{p}p$ total cross sections 
from accelerators and cosmic-ray experiments. From that plot, it is clear that
the mathematical description of the increase of the total cross
sections at the highest energies is an open problem. As we shall
discuss, the
study of the effects of the discrepant points at the highest
energies is one of our goals.
Figure 2 shows the typical diffractive pattern that characterizes
the differential cross section. We note that the data cover the
region corresponding to 10 decades. 
In Figure 3 it is displayed the slope parameter from
$pp$ and $\bar{p}p$ scattering as function of
the energy and determined in the region of small momentum transfer.
In what follows we shall refer to these three figures as
indicative of the empirical behavior of the quantities involved.

\begin{figure}
\begin{center}
\includegraphics[width=9.0cm,height=9.0cm]{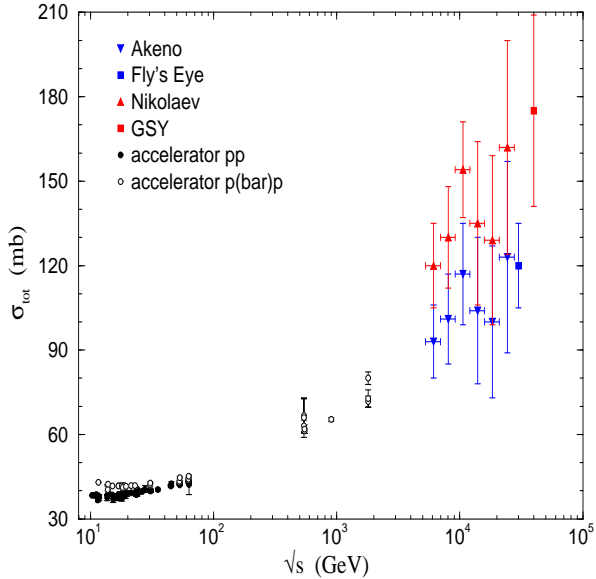}
\caption{Total cross sections on $pp$ and $\bar{p}p$ 
from accelerator and cosmic-ray experiments (for complete list
of references and tables see \cite{alm03}).}
\end{center}
\end{figure}

\begin{figure}
\begin{center}
\includegraphics[width=8.0cm,height=7.0cm]{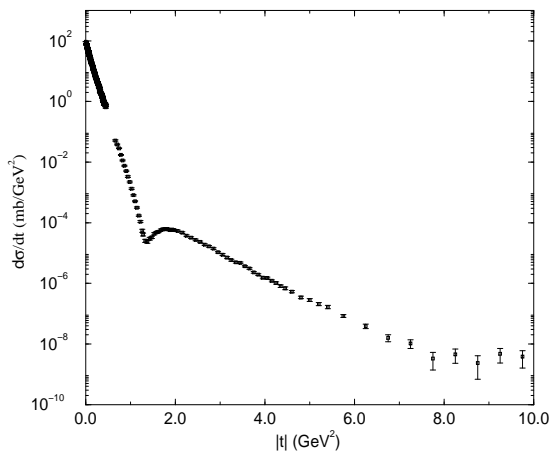}
\caption{Differential cross section data and the diffractive pattern
from $pp$ elastic scattering at $\sqrt s =$ 52.8 GeV.}
\end{center}
\end{figure}

\begin{figure}
\begin{center}
\includegraphics[width=6.cm,height=6cm]{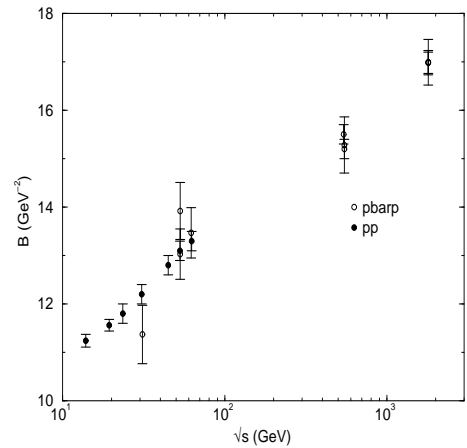}
\caption{The slope parameter as function of the energy and determined
in the interval 0.01 $ < |t| < $ 0.20 GeV$^2$.}
\label{fig5}
\end{center}
\end{figure}

\subsection{Principles, theorems and high-energy bounds}

For our purposes, we recall some principles and theorems from
axiomatic quantum field theories \cite{teo}. The basic Principles are:
Lorentz Invariance, Unitarity (related with the conservation of
probability), Analyticity (related to causality) and Crossing
(connecting particle-particle and particle-antiparticle
interactions).
Analyticity and crossing allow the connections between real and imaginary
parts of the scattering amplitude by means of dispersion relations.

Several rigorous theorems and bounds may be deduced from the
basic Principles and axiomatic quantum field theory. Among them,
the Froissart-Martin bound concerns the increase of the total cross
section stating that

\begin{eqnarray}
\sigma_{tot} \leq C \log^2 \frac{s}{s_0}
\qquad {\rm as} \qquad
s \rightarrow \infty. 
\end{eqnarray}

The Pomeranchuk Theorem treats the difference between
cross sections for particle-particle ($ab$) and particle-antiparticle
scattering ($a\bar{b}$). The original form was deduced when it was
believed that the cross section decreased to a constant value,
and in this case
$\sigma_{tot}^{ab} = \sigma_{tot}^{a\overline{b}}$
as $s \rightarrow \infty  $. After the discovery of the rising of
the cross section, Grunberg and Truong obtained the
generalized or revised form of the Pomeranchuk Theorem, stating
that 

\begin{eqnarray} 
\frac{\sigma_{tot}^{ab} - \sigma_{tot}^{a\bar{b}}}
{\sigma_{tot}^{ab} + \sigma_{tot}^{a\bar{b}}} \rightarrow 0
\quad {\rm or} \quad
\frac{\sigma_{tot}^{ab}} {\sigma_{tot}^{a\overline{b}}} \rightarrow 1
\quad {\rm as} \quad
s \rightarrow \infty, \nonumber 
\end{eqnarray}
and this means that, if the Froissart-Martin bound is reached, then

\begin{eqnarray} 
\Delta \sigma \equiv \sigma_{tot}^{a\overline{b}} - \sigma_{tot}^{ab}
\leq C \frac{ \sigma_{tot}^{a\overline{b}} + \sigma_{tot}^{ab}}{\log s} 
 \leq C \log s.
\end{eqnarray}

By expressing the cross sections in terms of crossing even ($+$)
and odd ($-$) contributions,

\begin{eqnarray} 
\sigma_{\pm}(s) = \frac{\sigma_{tot}^{ab} \pm 
\sigma_{tot}^{a\bar{b}}}{2} , \nonumber
\end{eqnarray}
we have
$|\Delta \sigma| = | \sigma_{tot}^{a\overline{b}} - 
\sigma_{tot}^{ab} | = 2\sigma_{-}$ .
Therefore,
$\Delta \sigma \equiv \sigma_{tot}^{a\overline{b}} - \sigma_{tot}^{ab}
\rightarrow 0$
if  and  only  if $\sigma_{-} 
\rightarrow 0$. This possible odd contribution is named Odderon
and the case of  even dominance at asymptotic energies is
associated with the Pomeron.

\subsection{Basic pictures}

Nearly all the phenomenological models, able to describe
the experimental data on elastic hadron scattering, are based
on the Optical/Geometrical Picture (s-channel) and/or 
the Exchange Picture (t-channel).
The corresponding formulas may be obtained from the
Partial Waves representation of the scattering amplitude,

\begin{eqnarray}
F(k,\theta) = \frac{i}{2k} \sum_{l=0}^{\infty} (2l + 1)
\left[ 1 -  e^{2i\delta_l} \right] P_l(\cos \theta), \nonumber
\end{eqnarray}
where $\delta_l$ is the phase shift. In what follows, we outline
the main steps and formulas in both pictures.

\subsubsection{Optical/Geometrical Picture}

From the partial wave representation, 
one considers the
high-energy limit and the semi-classical approximation, so that 
the discrete angular momentum $l$ may be replaced by
the continuum impact parameter $b$,
\begin{eqnarray}
l = kb - \frac{1}{2}. \nonumber
\end{eqnarray}
In turn, the discrete
phase shifts $\delta_l$ are replaced by the continuum eikonal
function of $b$ and $s$, $\chi(s,b)$ and 

\begin{eqnarray}
\sum_{l=0}^{\infty}...\ \ \rightarrow \ \ \int_{0}^{\infty} db...
\nonumber
\end{eqnarray}

The scattering amplitude 
in this \textit{Eikonal
Representation}, with azimuthal symmetry assumed, reads

\begin{eqnarray}
F(s, q) = ik \int_{0}^{\infty} bdb J_{0}(qb) 
[1 - e^{i\chi(s,b)}].
\end{eqnarray}

The quantity

\begin{eqnarray}
1 - e^{i\chi(s,b)} \equiv \Gamma (s,b) 
\end{eqnarray}
is named Profile function. From Unitarity this function is related to 
the probability that an inelastic event takes place at $b$ and $s$, 
the Inelastic Overlap function:

\begin{eqnarray}
G_{inel}(s,b) = | \Gamma(s,b) |^2 - 2Re \Gamma(s,b). 
\end{eqnarray}

Since in the Eikonal representation

\begin{eqnarray}
\ G_{inel}(s,b) = 1 - e^{-2 Im \chi(s,b)}, 
\end{eqnarray}
for
$Im\ \chi(s,b) \geq 0$ we have
$G_{inel}(s,b) \leq 1$, which implies in an automatically unitarized
representation.

\subsubsection{Exchange Picture}

In this picture, from the partial wave representation, one considers
the analytic continuation of the amplitude to complex angular momentum.
In the asymptotic limit ($s \rightarrow \infty$) and with symmetry
connecting the crossed channels
one arrives at the Watson-Sommerfeld-Gribov-Regge
representation for the scattering amplitude, expressed as a sum
over the poles of the amplitude (the Regge poles), as outlined in what 
follows. 

As it is known at high energies the number of partial waves is large,
and one way to circumvent that is to transform the sum of partial
waves into a complex integral, and then use the residues theorem
to obtain a new sum, but involving only the number of residues:

\begin{eqnarray}
\sum_{l=0}^{\infty} ... 
\rightarrow   \oint_{C} g(l) dl 
\rightarrow  \sum_{m=0}^{N} \left. Res\ g(l) \right|_{l =
l_{m}}. \nonumber
\end{eqnarray}

Detailed calculation allows one to obtain the following representation
for the scattering amplitude,

\begin{eqnarray}
F(k, \theta) = \sum_{i=1}^{N} 
\frac{\beta_i(k)P_{\alpha_{i}(k)}(-\cos \theta)}{
\sin \pi \alpha_i(k)} + BI(k, \theta), \nonumber
\end{eqnarray}
where $BI(k,\theta)$ is called the Background integral. 
By considering the high-energy limit (then $BI \rightarrow 0$) and crossing 
(exchange four-momenta $p \rightarrow \quad
\Leftrightarrow \quad
\leftarrow -\bar{p}$)
we can replace the crossing channel variable
($\bar{\theta} \leftrightarrow s$)

\begin{eqnarray}
\cos \bar{\theta} = 1 - \frac{2s}{4m^2 - t} \quad
\rightarrow \quad \propto
- s \quad {\rm as} \quad s \rightarrow \infty, \nonumber
\end{eqnarray}
also,

\begin{eqnarray}
P_l(x) \rightarrow \left[\frac{2^l \Gamma(l + 1/2)}
{\sqrt \pi \Gamma(l + 1)}\right] x^l \quad {\rm for} \quad
x \rightarrow \pm \infty, \nonumber
\end{eqnarray}
and grouping all the $s$-independent quantities in a function 
${\cal K}(t)$ we have

\begin{eqnarray}
P_{\alpha(t)}(-\cos \bar{\theta}) = {\cal K}(t)s^{\alpha(t)}
\quad {\rm for} \quad s \rightarrow \infty. \nonumber
\end{eqnarray}

Rearranging the terms we arrive at
a descending asymptotic series in powers of $s$, with leading 
contribution:

\begin{eqnarray}
F(s,t) = \gamma(t) \xi(t) s^{\alpha(t)},
\end{eqnarray}
where $\gamma(t)$ is the residue function, 
$\xi(t)$ the signature factor and
$\alpha(t) = \alpha(0) + \alpha' t$ the trajectory function.
This last function connects the spin and masses through the
Chew-Frautschi plot, as exemplified in Fig. 4.
In this picture the interaction of the colliding particles
is basically interpreted in terms of exchanges of Regge
poles (also Regge cuts) and the Pomeron (an \textit{ad hoc} trajectory
with intercept nearly above 1). We note that, as constructed,
the exchange picture is intended for asymptotic energies.

\begin{figure}
\begin{center}
\includegraphics[width=8.0cm,height=7.0cm]{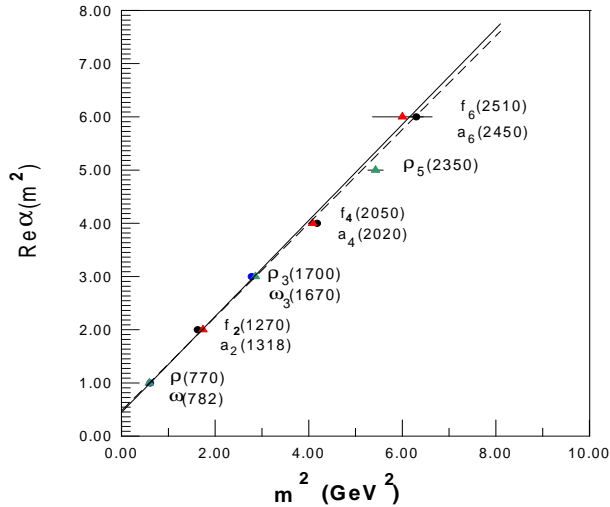}
\caption{The Chew-Frautschi plot for some mesons and resonances.}
\end{center}
\end{figure}

\section{Analytic approach}

The Analytic Approach for elastic hadron-hadron scattering is based
on general principles and theorems from Quantum Field Theory.
It is
characterized by analytical parametrizations for
the imaginary part of the forward amplitude, together with the use of 
dispersion relation
techniques. The central point is the simultaneous investigation
of the total cross section (imaginary part
of the scattering amplitude, Eq. (2)) and the
$\rho$ parameter (connected with the real part of the
amplitude, Eq. (3)).

For particle-particle and particle-antiparticle interactions, 
dispersion relations  are
consequences of the principles of Analyticity and Crossing. 
In this context, they correlate real and imaginary
parts of crossing even ($+$) and odd ($-$) amplitudes,
which in turn are expressed in terms of the scattering amplitudes for
a given process and its crossed channel, for example, $ a + b $ and
$a + \bar{b} $:

\begin{equation}
F_{ab} = F_{+} + F_{-}, \qquad
F_{a\bar{b}} = F_{+} - F_{-}.
\end{equation}

At high energies, the standard singly subtracted integral dispersion 
relations,
with poles removed, are given by

\begin{equation}
\re F_{+}(s)= K + \frac{2s^{2}}{\pi}P\!\!\!\int_{s_{0}}^{+\infty}
\!\!\!\ d s'
\frac{1}{s'(s'^{2}-s^{2})}\im F_{+}(s')
\label{eq:4}
\end{equation}
and

\begin{eqnarray}
\re F_{-}(s)=  \frac{2s}{\pi}P\!\!\!\int_{s_{0}}^{+\infty} \!\!\!
\ d s'
\frac{1}{(s'^{2}-s^{2})}\im F_{-}(s'),
\label{eq:5}
\end{eqnarray}
where $K$ is the subtraction constant and, for $pp$ and $\bar{p}p$
scattering, $s_0=2m^2\sim 1.8$ GeV$^2$.

In this section we review some results obtained through this approach.
We start with the replacement of the above integral forms by derivative
operators (Derivative Dispersion Relations) and then we discuss
the use of analytic
models (Reggeons, Pomeron, Odderon) for parametrizations involving the total 
cross section
and the $\rho$ parameter, the determination of bounds for
the soft Pomeron intercept, and the practical role of the subtraction constant.

In what follows we are mainly concerned with the $pp$ and $\bar{p}p$ 
elastic scattering,
since for particle and antiparticle interactions they correspond to
the highest energy
interval with available data and are the only set including the cosmic-ray 
information on total cross sections ($pp$ scattering).
As commented before, the experimental data available on the
total
cross sections (Figure 1) are characterized by discrepant experimental
information at the highest energies, and one of our aims is to
investigate the effects of these discrepancies in the context
of the analytic models. This concern permeates all the
discussion in this Section.

\subsection{Derivative Dispertion Relations}

The use of dispersion relations in the investigation of scattering
amplitudes may be traced back to the end of fifties, when they were
introduced in the form of {\em Integral} Dispersion Relations (IDR). 
Despite the
important results that have been obtained since then, one limitation of
the integral forms is their non-local character:
in order to obtain the real part of the amplitude, the imaginary part
must be known for all values of the energy. Moreover, the class of
functions that allows analytical integration is limited.

In the last years, we have investigated the applicability of {\em Derivative}
Dispersion Relations (DDR) 
in place of integral forms \cite{mmp,alm01,almhadron02,alm03,lm03,am03}.
In Reference  \cite{am03} we present a 
recent review on different results 
and statements
related to this replacement,
and a discussion 
connecting these different aspects with the corresponding assumptions and
classes of functions considered in each case.

In particular, we have shown that for the class of functions which
are entire in the logarithm of the energy (as is the case of analytic
models at high energies)
it is possible to expand the integrand in the above formulas
and by considering a high-energy approximation, represented
by $s_0=2m^2 \rightarrow 0$, to integrate term by term. 
In that case, as demonstrated in detail in
\cite{am03}, the derivative dispersion relations with
one subtraction reads

\begin{equation}
\frac{\re F_+(s)}{s}= \frac{K}{s} +
\tan\left[\frac{\pi}{2}
\frac{\mathrm{d}}{\mathrm{d}\ln s}\right]\frac{\im F_+(s)}{s},
\end{equation}

\begin{equation}
\frac{\re F_-(s)}{s}=
\tan\left[\frac{\pi}{2}
\left(1 + \frac{\mathrm{d}}{\mathrm{d}\ln s}\right)\right]
\frac{\im F_-(s)}{s},
\label{eq:11}
\end{equation}
where the series expansion is implicit in the tangent operator.
From this deduction one arrives to three formal results: (1) the
subtraction constant is preserved when the IDR are replaced by DDR and,
therefore, in principle, can not be disregarded in fit procedures;
(2) except for the subtraction constant, the DDR with entire functions
in the logarithm of the energy do not depend on any
additional free parameter; (3) the only approximation involved in the
replacement concerns the lower limit in the IDR (13-14), namely
$s_0 = 2m^2 \rightarrow 0$, which represents a high-energy 
approximation. 
In the next two subsections we discuss some uses of the DDR
with analytical models,
and in the third subsection we return to the replacement of IDR
by DDR, investigating the important role of the subtraction constant
from a practical point of view.

\subsection{Basic Models}

In this Subsection we make use of two basic and well
known parametrizations for the
total cross sections and investigate the effects 
of the discrepancies in the experimental information
from cosmic-ray experiments.

\subsubsection{Ensembles}

In the cosmic-ray region, $6\ \textrm{TeV} < \sqrt s \leq 40$ TeV, the 
discrepancies 
on the total cross section information are
due to both experimental and theoretical uncertainties in the 
determination of
$\sigma_{tot}^{pp}$ from p-air cross sections. The situation has been 
recently 
reviewed in detail in \cite{alm03}, where a complete list of references, 
numerical 
tables and discussions are presented. 

From Fig. 1 we see that, despite the large error bars in
the cosmic-ray region, we can
identify two distinct sets of estimations: one corresponding to the results 
by
the Fly's Eye Collaboration (Fly's Eye) together with those by the Akeno
Collaboration (Akeno); the other set associated with the results by Gaisser, 
Sukhatme, 
and Yodh
(GSY) together with with those by Nikolaev (Nikolaev).
Taken separately these two sets suggest different
scenarios for the increase of the total cross section,
as previously discussed in \cite{alm01,mm97,lm02}.

Based on these considerations, it is important to investigate the 
behavior of the
total cross section by taking into account  the discrepancies that
characterize the cosmic ray information.  
To this end, in \cite{alm03} we have considered \textit{two ensembles 
of data and
experimental information}, as follows:

\vspace{0.3cm}

$\bullet$ Ensemble I:
$\bar{p}p$ and $pp$ accelerator data + Akeno +
 Fly's Eye;

\vspace{0.3cm}

$\bullet$ Ensemble II:
$\bar{p}p$ and $pp$ accelerator data + Nikolaev + 
GSY.

\vspace{0.3cm}

To some extent, ensemble I represents a kind of high-energy standard 
picture and ensemble II a nonstandard one.

\subsubsection{Analytic Models}

With analytical parametrizations for $pp$/$\bar{p}p$ total cross
sections, the connections with the $\rho$ parameter, Eq. (3),
are obtained by defining the associated crossing even and odd
quantities,

\begin{eqnarray}  
\sigma_{\pm}(s) = \frac{\sigma_{tot}^{pp} \pm 
\sigma_{tot}^{\bar{p}p}}{2},
\end{eqnarray}
using the high-energy normalization for the Optical Theorem,

 \begin{eqnarray}
\sigma_{tot}(s) 
\sim \frac{\im F(s,0)}{s}, 
\end{eqnarray}
and the DDR given by Eqs. (15) and (16).

In \cite{alm03} we have considered two different parametrizations for 
the total cross
sections, one introduced by Donnachie and Landshoff \cite{dl} and  
other by
Kang and Nicolescu \cite{kn}.
The main difference concerns the asymptotic limits, which allow  the
dominance of an even amplitude (Pomeron) or 
the odd amplitude (Odderon), respectively. In this way, we may contrast
these possibilities with the standard and non-standard pictures
represented by Ensembles I and II.

The \textit{Donnachie-Landshoff} (DL) parametrization for the total 
cross sections
is expressed by 

\begin{eqnarray}
\sigma_{tot}^{pp} (s) = X s^{\epsilon} + Y s^{- \eta},  
\qquad
\sigma_{tot}^{\bar{p}p} (s) = X s^{\epsilon} + Z s^{- \eta}, 
\end{eqnarray} 
where the first contribution is associated with a single Pomeron 
exchange (universal) and the second one with Reggeon exchange.
With the procedure explained above, we obtain the analytical
connections with the $\rho$ parameter for $pp$ and $\bar{p}p$
scattering:

\begin{eqnarray}
&&\rho ^{pp}(s)  \sigma_{tot}^{pp}(s) = 
\frac{K}{s} +
\left[X\tan\left(\frac{\pi\epsilon}{2}\right)\right]s^{\epsilon}  
\nonumber \\
&+&  
\left[\frac{(Y-Z)}{2}\cot\left(\frac{\pi\eta}{2}\right) -
\frac{(Y+Z)}{2}\tan\left(\frac{\pi\eta}{2}\right)\right]s^{- \eta}, 
\nonumber
\end{eqnarray}

\begin{eqnarray}
&&\rho ^{\overline{p}p}(s) \sigma_{tot}^{\overline{p}p}(s) =
 \frac{K}{s} +
\left[X\tan\left(\frac{\pi\epsilon}{2}\right)\right]s^{\epsilon} 
\nonumber \\
&+&
\left[\frac{(Z-Y)}{2}\cot\left(\frac{\pi\eta}{2}\right) -
\frac{(Y+Z)}{2}\tan\left(\frac{\pi\eta}{2}\right)\right]s^{- \eta}.
\nonumber
\end{eqnarray}

From the above formulas, since $\eta > 0$, this model predicts that,
asymptotically ($s \rightarrow \infty$),

\begin{eqnarray}
&\Delta \sigma& = \ \sigma^{\overline{p}p}_{tot}(s) - 
\sigma^{pp}_{tot}(s) 
\rightarrow 0, \nonumber \\
&\Delta \rho& = \ \rho^{\bar{p}p}(s) -  \rho^{pp}(s) \rightarrow 0.
\nonumber
\end{eqnarray}

The parametrization for the total cross sections introduced by 
\textit{Kang and Nicolescu} (KN), under the 
hypothesis of the Odderon, is given by 

\begin{eqnarray}
\sigma^{pp}_{tot}(s)= A_{1} + B_{1} \ln s  +
k \ln^2 s, \nonumber
\end{eqnarray}

\begin{eqnarray}
\sigma^{\overline{p}p}_{tot}(s)= A_{2} + B_{2} \ln s 
+ k \ln^2 s + \frac{2R}{s^{1/2}}, \nonumber
\end{eqnarray}
and the connections with $\rho$ read

\begin{eqnarray}
&&\rho ^{pp}(s) \sigma_{tot}^{pp}(s)= 
\frac{K}{s} + \frac{\pi}{2}\left(\frac{B_1 + B_2}{2}\right) 
\nonumber \\
&+& 
\left(\pi
k + \frac{A_2 - A_1}{\pi}\right) \ln s + \left(\frac{B_2 - B_1}{2\pi} 
\right)
\ln ^2 s  - \frac{2R}{s^{1/2}}, \nonumber
\end{eqnarray}

\begin{eqnarray}
&&\rho ^{\overline{p}p}(s) \sigma_{tot}^{\overline{p}p}(s) = 
\frac{K}{s} + \frac{\pi}{2}\left(\frac{B_1 +
B_2}{2}\right) \nonumber \\ 
&+& \left(\pi k - \frac{A_2 - A_1}{\pi}\right)\ln s -
\left(\frac{B_2 - B_1}{2\pi} \right) \ln ^2 s  \nonumber. 
\end{eqnarray}

Differently from the previous case, this model predicts that the difference
 between the two 
cross sections is given by

\begin{eqnarray}
&\Delta \sigma&  =
(A_2 - A_1) + (B_2 - B_1)\ln s + 2Rs^{-1/2} \nonumber \\
&\rightarrow& \Delta A + 
\Delta B \ln s \quad {\rm(asymptotically)}, \nonumber
\end{eqnarray}
so that, if $\Delta A
\not= 0$ and/or  $\Delta B \not= 0$, the total cross section difference may 
increase and 
$\sigma_{tot}^{pp}$ may even become greater than $\sigma_{tot}^{\bar{p}p}$, 
depending on the
values and signs of $\Delta A$ and $\Delta B$, which is formally in 
agreement with
the theorems of Sec. II.B.
Moreover, if $\Delta A$ and $\Delta B$ are sufficiently small, so that
we may replace $\sigma_{tot}^{\bar{p}p} \approx \sigma_{tot}^{pp} \equiv 
\sigma_{tot}(s)$, then, 
asymptotically,

\begin{eqnarray}
\Delta \rho = \rho ^{\overline{p}p}  - \rho ^{pp} \sim
- \frac{1}{\pi \sigma_{tot}(s)} \left\{ \Delta A \ln s + \Delta B \ln^2 s 
\right\}. \nonumber
\end{eqnarray}
This means that, depending on the fit results, there may be a change of 
sign in
$\Delta \rho$, with $\rho ^{pp}$ becoming greater than 
$\rho ^{\overline{p}p}$ at some
finite energy. Therefore, the case of a crossing either in $\sigma_{tot}$
or $\rho$ is a sign of the odderon contribution in the imaginary or
real part of the amplitude, respectively.

\subsubsection{Fits and Results}

We have performed 16 different fits through the program CERN-MINUIT. In these 
fits we have used both
ensembles I and II  and both the DL and KN models. 
For each of these four possibilities we have performed global and 
individual
fits to $\sigma_{tot}$ and $\rho$ and, in each case, we either considered the 
subtraction
constant $K$ as a free fit parameter, or assumed $K = 0$. 

All the results are presented and discussed in detail in Ref. \cite{alm03}. 
Our main conclusions are the following: 
(1) Despite the small influence from different cosmic-ray estimations, the
results allow to extract an upper bound for the soft Pomeron 
intercept: $1 + \epsilon = 1.094$;
(2) although global fits present good statistical results, in general, this
procedure constraints the rise of  $\sigma_{tot}$; 
(3) the subtraction constant as a free
parameter affects the fit results at both low and high energies; 
(4) independently of the 
cosmic-ray information used and the subtraction constant, global fits with 
the Odderon 
parametrization predict that, above $\sqrt s \approx 70$ GeV, $\rho_{pp}(s)$ 
becomes greater 
than $\rho_{\bar{p}p}(s)$, and this result is in complete
agreement with all the data presently available. That result is displayed
in Fig. 5 and  we can infer
$\rho_{pp} = 0.134\ \pm \ 0.005$ at $\sqrt s = 200$ GeV and $0.151\ \pm 
\ 0.007$ at $500$ GeV (BNL RHIC energies).

\begin{figure}
\begin{center}
\includegraphics[width=8.0cm,height=7.0cm]{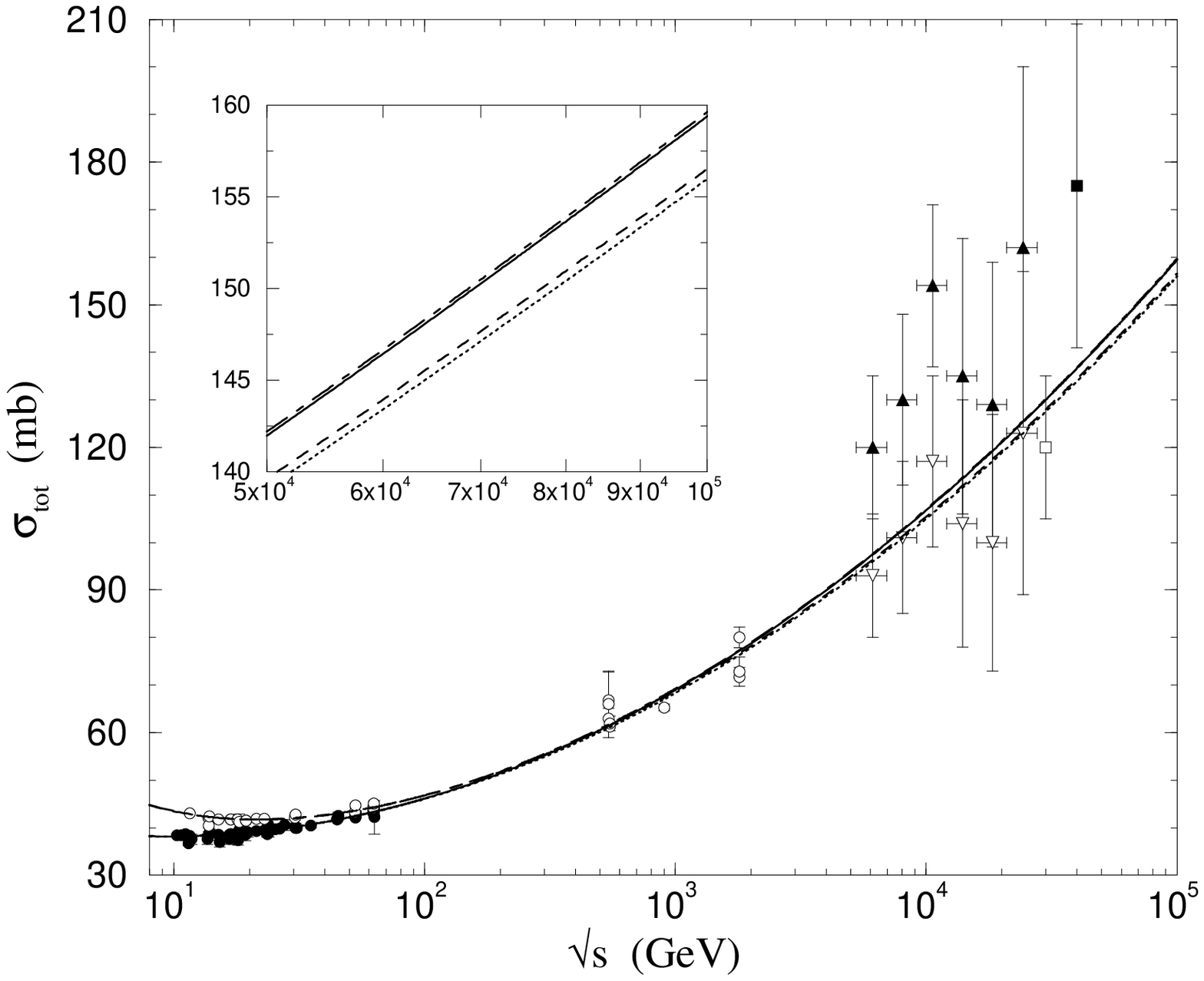}
\end{center}
\begin{center}
\includegraphics[width=8.0cm,height=7.0cm]{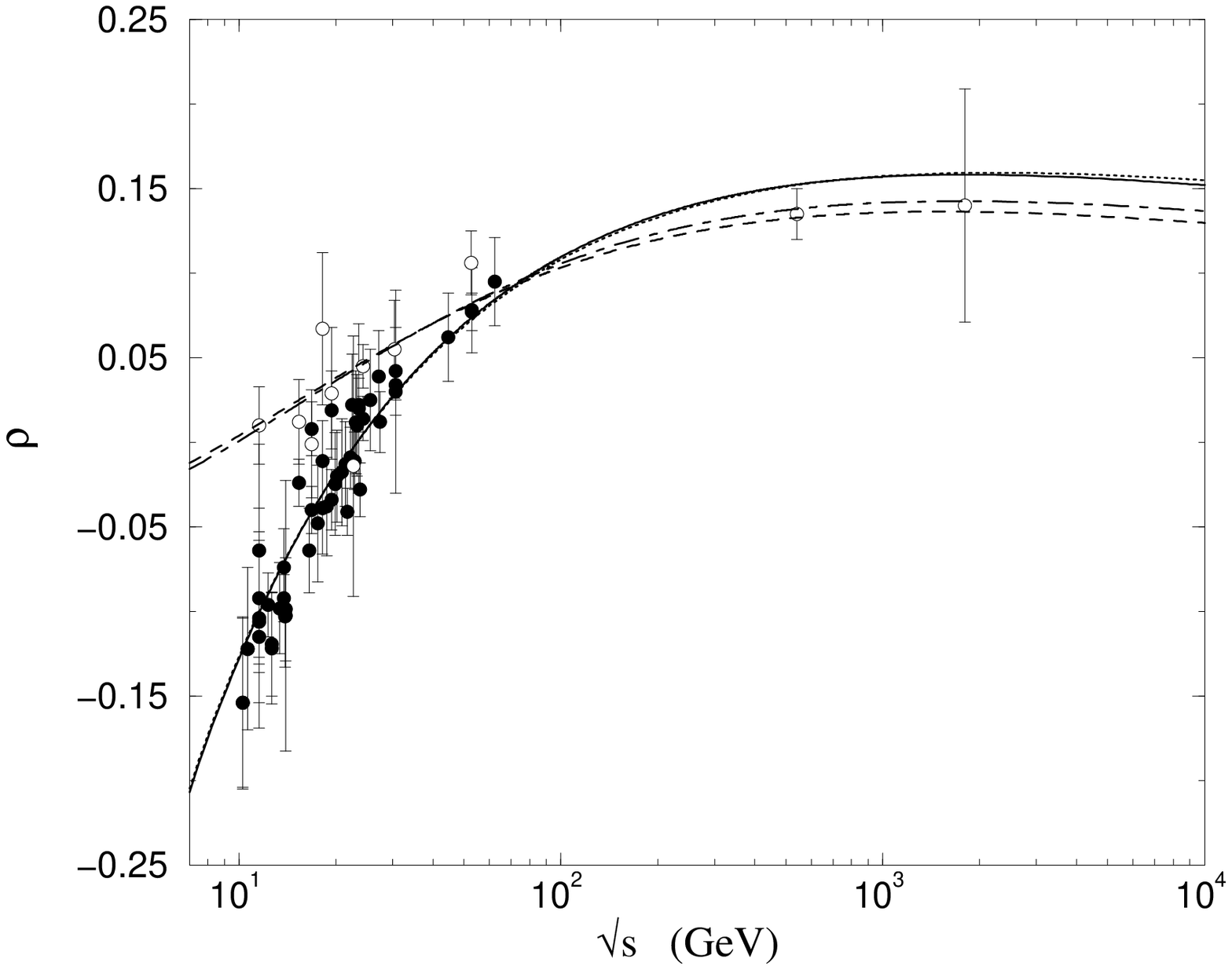}
\caption{ Simultaneous fits to $\sigma_{tot}(s)$ and $\rho(s)$ through the KN
parametrization with $K=0$ and ensembles I (dotted curves for $pp$ and dashed 
for $\bar{p}p$) 
and II (solid curves for $pp$ and dot-dashed for $\bar{p}p$) \cite{alm03}.}
\end{center}
\end{figure}

\subsection{Non-degenerate Meson Trajectories}

The DL parametrization referred above
assumes degeneracies between the secondary reggeons, imposing a common 
intercept for the $C=+1$ ($a_2, f_2$) and the $C=-1$ ($\omega,\rho$)
trajectories (see Fig. 4).
More recently, analysis treating global fits to
$\sigma_{tot}$ and $\rho$ have indicated that the best results are 
obtained with 
non-degenerate meson trajectories. In this case the forward
scattering amplitude is decomposed into three reggeon exchanges,
$
F(s) = F_{\tt I\!P}(s) + F_{a_2/f_2}(s) + 
\tau F_{\omega/\rho}(s), 
$
where the first term represents the exchange of a single soft Pomeron, 
the other
two the secondary Reggeons and $\tau = + 1$ ($- 1$) for $pp$ ($\bar{p}p$)
amplitudes. Using the notation $\alpha_{\tt I\!P}(0) 
= 1+\epsilon$, $\alpha_{+}(0) = 1 -\eta_{+}$ and $\alpha_{-}(0) = 
1 -\eta_{-}$ 
for the intercepts of the Pomeron and the $C=+1$ and $C=-1$ 
trajectories, respectively, the total cross sections, Eq. (18), for $pp$ 
and $\bar{p}p$ 
interactions are 
written as

\begin{eqnarray}
\sigma_{tot}(s) = X s^{\epsilon} + Y_{+}\, s^{-\eta_{+}} + \tau Y_{-}\, 
s^{-\eta_{-}}
\end{eqnarray}
and the connection with the $\rho$ parameter by means of DDR is similar 
to that
displayed in the last subsection. 

Making use of this parametrization, in this section we  present the 
determination of
extrema bounds for the Pomeron intercept \cite{lm03} and a practical
analysis on the replacement of IDR by DDR together with a discussion
on the role of the subtraction constant \cite{am03}.

\subsubsection{Extrema Bounds for the Pomeron Intercept}

In order to analyze the extrema effects in the soft Pomeron
intercept due to discrepancies in
the experimental data, we performed a detailed analysis 
including the highest and the lowest values of the total cross
section from both accelerators and cosmic-ray experiments.

As it is well known, in the accelerator region, the conflict concerns 
the results
for $\sigma_{tot}^{\bar{p}p}$ at $\sqrt s = 1.8$ TeV reported by the  
CDF Collaboration and those reported by the E710  and the  
E811 Collaborations (Fig. 1).
In the cosmic-ray region, as we have discussed, the highest predictions for 
$\sigma_{tot}^{pp}$
concern the result by Gaisser, Sukhatme, and Yodh together with those by
Nikolaev. In order to treat the lowest estimations in the cosmic-ray
region, we consider
the results obtained by Block, Halzen, and Stanev (BHS), by means of
a QCD-inspired model. As discussed in \cite{alm03}, the reason
for this choice is that, although the extracted
$\sigma_{tot}^{pp}(s)$  shows agreement with the 
Akeno results, it is about $17$ mb below the Fly's Eye value at $30$ TeV
and therefore may be considered as a extreme lower estimate. All the
numerical tables and references can be found in \cite{alm03}.

In this case we have considered the following ensembles of experimental
information. 
First we only consider 
accelerator data in two ensembles with the following notation:

\vspace{0.3cm}

$\bullet$ Ensemble I: $\sigma^{pp}_{tot}$ and $\sigma^{\bar{p}p}_{tot}$ 
data ($10 \le \sqrt{s} \le 
900$ GeV) + CDF datum ($\sqrt{s} = 1.8$ TeV);

\vspace{0.3cm}

$\bullet$ Ensemble II
: $\sigma^{pp}_{tot}$ and $\sigma^{\bar{p}p}_{tot}$ 
data ($10 \le \sqrt{s} \le 
900$ GeV) + E710/E811 data ($\sqrt{s} = 1.8$ TeV).

\vspace{0.3cm}
 
Ensemble I represents the faster increase scenario for the rise of 
$\sigma_{tot}$ from
accelerator data and ensemble II the slowest one. These ensembles 
are then combined with
the highest and lowest estimations for $\sigma_{tot}^{pp}$ from 
cosmic-ray experiments,
namely, the Nikolaev and the Gaisser, Sukhatme, and Yodh (NGSY) 
results and 
the Block, Halzen, 
and Stanev (BHS) results, respectively. These new ensembles are 
denoted by 

\vspace{0.3cm}

$\bullet$ I + NGSY

\vspace{0.3cm}

$\bullet$ II + BHS

\vspace{0.3cm}

As in the previous analysis, we have considered both individual 
fits to $\sigma_{tot}$, 
and 
simultaneous fits to $\sigma_{tot}$ and $\rho$, either in the case 
where the subtraction 
constant is considered as a free fit parameter or assuming $K = 0$. 

From this analysis, in the case of only accelerator data,
we could infer the following upper and lower
values for the Pomeron intercept: $\alpha_{\tt I\!P}(0)=1.098\pm 0.004$
(global fits to ensemble I, with $K = 0$)
and $\alpha_{\tt I\!P}(0)=1.085\pm 0.004$
(individual fit to $\sigma_{tot}$ from ensemble II), 
with bounds $1.102$ and $1.081$, respectively. 
Adding the cosmic-ray information, we inferred
the following upper and lower values:
$\alpha_{\tt I\!P}(0)=1.104\pm 0.005$
(individual fit to $\sigma_{tot}$ from ensemble I + NGSY)
and $\alpha_{\tt I\!P}(0)=1.085\pm 0.003$
(global fits to ensemble II + BHS and $K$ as a free fit parameter
or individual fit to $\sigma_{tot}$ from this ensemble), 
with bounds $1.109$ and $1.082$,
respectively. Therefore we may infer the following \textit{extrema} 
bounds
for the soft Pomeron intercept:

\begin{eqnarray}
\alpha^{upper}_{\tt I\!P}(0) = 1.109,
\qquad
\alpha^{lower}_{\tt I\!P}(0) = 1.081. \nonumber
\end{eqnarray}
Figure 6 shows the total cross sections with parametrization
(20) and the above extrema bounds,
together with the experimental information available.

\begin{figure}
\begin{center}
\includegraphics[width=8.0cm,height=7cm]{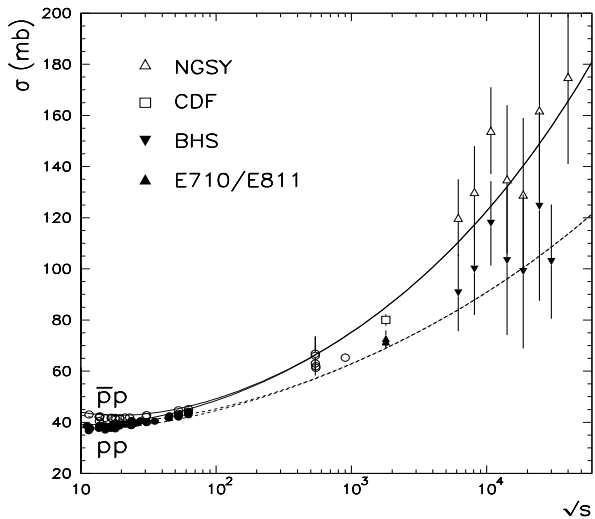}
\end{center}
\caption{Fastest and 
slowest increase scenarios for the rise of the total cross section
through parametrization (20) and
allowed by the experimental information available: fits to ensembles
I + NGSY (solid) and II (dashed) \cite{lmm03}.}
\end{figure}

Extensions of these \emph{extrema bounds} for the pomeron intercept to 
meson-p, gamma-p and gamma-gamma scattering have been
discussed in \cite{lmm03}. By means of global fits to total cross section
data it is shown that these bounds are in 
agreement
with the bulk of experimental data presently available, and  
extrapolations
to higher energies indicate different behaviors for the rise of the
total cross sections.

We have also obtained new \emph{constrained bounds} for the Pomeron
intercept from spectroscopy data
(Chew-Frautschi plots) and have extended the analysis to baryon-p, meson-p, 
baryon-n, meson-n,
gamma-p and gamma-gamma scattering \cite{lmm-npa04}. It is also
presented tests on factorization
and quark counting rules with both extrema and constrained bounds
(asymptotic energy region). In particular, at 14 TeV (CERN LHC)
the extrema and constrained bounds allow to infer
$\sigma_{tot} = 114 \pm 25$ mb and $105 \pm 10$ mb, respectively.
\cite{lmm-npa04}.

\subsubsection{IDR, DDR and the Subtraction Constant}

As commented before, we have shown in Ref. \cite{am03} that for entire 
functions in the logarithm of the energy
the only approximation involved in the replacement of IDR by
DDR concerns the lower limit $s_0$ in the IDR: the high-energy
condition is reached by assuming that $s_0 = 2m^2 \rightarrow 0$
in Eqs. (13-14).
In that paper we have investigated the practical applicability of the
 DDR and IDR in the
context of the Pomeron-reggeon parametrizations, with both degenerate 
and
non-degenerate higher meson trajectories. By means of global fits to
$\sigma_{tot}(s)$ and $\rho(s)$ data from $pp$ and $\bar{p}p$ scattering,
we have tested all the 16 important variants that could affect the fit
results, namely the number of secondary reggeons, energy cutoff
(5 and 10 GeV), effects of
the high-energy approximation connected with the subtraction constant 
and the 
analytic
approach using both DDR and IDR with fixed $s_0$. Our results led to 
the conclusion that
the high-energy approximation and the subtraction constant affect the
fit results at both low and high energies. This effect is a consequence 
of the 
fit procedure, associated with the strong correlation among the free 
parameters.

A striking novel result concerns the practical role of the subtraction
constant. We have shown that, with the Pomeron-reggeon parametrizations,
once the subtraction constant is used as a free fit parameter, the 
results 
obtained with the DDR and with the IDR (with finite lower limit, 
$s_0 = 2m^2$)
are the same up to 3 significant figures in the fit parameters and
$\chi^2/DOF$. This conclusion, as we have shown, is independent of the
number of secondary reggeons (DL or extended parametrization) or the
energy cutoff ($\sqrt s$ = 5 or 10 GeV). In Table I we display the
fit results with the extended parametrization and cutoff at 10 GeV.

\begin{table}[h]
\begin{center}
\caption{Simultaneous fits to $\sigma_{tot}$ and $\rho$ 
through the extended parametrization, $\sqrt s_{\mathrm{min}} =$ 
10 GeV
(154 data points), with $K$ as a free
parameter  and using IDR with lower limit $s_0=2m^2$ and DDR
\cite{am03}.}
\label{tab:4}
\begin{tabular}{ccc}
\hline
           & IDR with $s_0=2m^2$  & DDR         \\
\hline
$X$        &   19.57  $\pm$ 0.79   & 19.58  $\pm$ 0.78      \\
$Y_+$      &   66.0   $\pm$ 6.7    & 66.0   $\pm$ 6.6       \\
$Y_-$      &   -29.2  $\pm$ 4.0    & -29.2  $\pm$ 4.0       \\
$\epsilon$ &   0.0897 $\pm$ 0.0033 & 0.0897 $\pm$ 0.0033  \\
$\eta_+$   &   0.380  $\pm$ 0.033  & 0.380  $\pm$ 0.033    \\
$\eta_-$   &   0.520  $\pm$ 0.025  & 0.520  $\pm$ 0.024   \\
$K$        &   -14    $\pm$ 48     & 104    $\pm$ 58      \\
$\chi^2/DOF$ &   1.10      & 1.10                 \\
\hline
\end{tabular}
\end{center}
\end{table}

\section{ Model Independent Analysis}

This kind of analysis is characterized by model independent parametrizations
of the experimental data involved and the extraction of empirical or
semi-empirical information that can contribute with the development
of phenomenological models and the underlying theory. In this
section we review some results we have obtained in the investigation
of $pp$ and $\bar{p}p$ differential cross section data (unconstrained
and constrained fits, as will be explained) and the correlations
between the experimental data on total cross section and the slope
parameter (Eq. (4)).

\subsection{Differential Cross Section}

Several authors have investigated elastic hadron scattering by means
of parametrizations for the scattering amplitude and fits to the 
differential
cross section data, Eq. (1). The extraction of the Profile, Eikonal and
Inelastic Overlap functions in the $b$-space (impact parameter) and, 
in some special
cases, the Eikonal in the $q^2$-space, has led to important and novel 
results related with geometrical aspects (radius, central
opacity), differences between charge distributions and hadronic matter
distributions, existence or not of eikonal zeros in the $q^2$-space and,
more recently, connections with pomerons, reggeons and nonperturbative
QCD aspects. In Ref. \cite{cmm03} we present a review and a critical 
discussion
on the main results concerning this kind of analysis and also a wide
list of references to outstanding works.

The basic input in all these analyses is the parametrization of the
scattering amplitude as a sum of exponentials in $q^2$ 
(as empirically suggested by the diffractive pattern shown in
Fig. 2) and fits
to the differential cross section data. This parametrization allows
analytical expressions for the Fourier transform of the amplitude,
providing also analytical expressions for the quantities of interest
in the $b$-space.

In the next two subsections we review the results we have obtained
by means of unconstrained fits (fit parameters completely free, without
extracted dependences on the energy) \cite{cmm03,cm}, and discuss some
research in course related to constrained fits (including dependences 
on the energy which are based on empirical information) \cite{acmmhadron04}.

\subsubsection{Unconstrained Fits and the Eikonal}

In the high energy region, $\sqrt s > $10 GeV,
differential cross section data are available at 
$\sqrt s =$ 13.8, 19.5, 23.5, 30.7, 44.7, 52.8 and 62.5 GeV for $pp$
scattering and at $\sqrt s =$ 13.8, 19.4,
 31, 53, 62, 546  and 
1800 GeV for $\bar{p}p$ scattering.
Data from $pp$ scattering also exists at
$\sqrt s =$ 27.5 GeV 
and 5.5 $\leq q^2 \leq$ 14.2 GeV$^2$ (but not on $\sigma_{tot}$ 
and $\rho$), and as we shall show, that
set plays a fundamental role in our analyse. See \cite{cmm03} for
a complete list of references.

As discussed in \cite{cmm03} two main problems are typical of
model independent analysis of the differential cross sections:

(1) Experimental data are available only over finite regions
of the momentum transfer  (which in general are small, 
$q^2 < $ 7 GeV$^2$) and the Fourier transform demands integration
from $q^2$ = 0 to infinity. This means that any fit is biased by 
extrapolations and although some extrapolated
curves may look unphysical, they can not be excluded on
mathematical grounds.

 (2) The exponential parametrization allows analytical determination
of the quantities in the $b$-space (profile, inelastic, eikonal
functions)
and also the statistical
uncertainties, by means of error propagation from the fit parameters. 
However,
in this case, the translation of the eikonal
from b-space to the $q^2$-space can not be analytically
performed and neither the error propagation (through standard 
procedures).
As a consequence, the unavoidable uncertainties from the fit 
extrapolations
can not, in principle, be taken into account.

In what follows we review a model independent approach able to
minimize the above two problems.

- \leftline{\textit{Fit Procedure}}

In order to treat problem (1) we have used the following 
procedure \cite{cm}.
Since it is known that for large $t$ the experimental data 
do not depend on the energy at 
$13.8$ GeV $\leq \sqrt s \leq 62$ GeV and that there exist
data at $\sqrt s = 27.5$ GeV in the region
$5.5$ GeV$^{2} \leq q^{2} \leq 14.2$ GeV$^{2}$, we have selected
two ensembles of 
$pp$ and $\bar{p}p$ differential cross section data:  

$\bullet$ Ensemble I: experimental data at each energy;

$\bullet$ Ensemble II: Ensemble I  + data at $\sqrt s = 27.5$ GeV.

For the scattering amplitude we have introduced the following
model independent
analytical parametrization for both real and imaginary parts:

\begin{eqnarray} 
F(s,q) = 
 \{ \mu  \sum_{j=1}^{2} \alpha_{j} e^{-\beta_{j} q^{2}}
\} + i
\{ \sum_{j=1}^{n} \alpha_{j} e^{-\beta_{j} q^{2}} \}, 
\end{eqnarray}

\begin{eqnarray}
\mu = \frac{\rho(s)}{\alpha_{1} + \alpha_{2}} \sum_{j=1}^{n} 
\alpha_{j}. 
\end{eqnarray} 

With the experimental $\rho$ value at each energy the
fits to the differential cross section data have been performed
through the CERN-MINUIT routine 
and the validity or not of ensemble II is checked by means of
the MINUIT output  and standard statistical
interpretation of the fit results (DOF, confidence levels). 

For $pp$ scattering 
we have found that the data at $\sqrt s =$ 13.8 GeV are
not compatible with ensemble II. In the case of $\bar{p}p$ scattering
none of the data sets are compatible with ensemble II.
Therefore, in what follows, ensemble II (data at
$\sqrt s =$ 27.5 GeV added) corresponds only to $pp$ scattering
at 6 energies: 19.5, 23.5, 30.7, 44.7, 52.8 and 62.5 GeV.

From the error matrix (variances and covariances),
$\chi^2/DOF$ and confidence intervals, we
infer the best values for the parameters and corresponding
errors $\Delta\alpha_{j}$, $\Delta\beta_{j}$. 
By means of standard error propagation,
the uncertainties in the free parameters, $\Delta \alpha_{j}$,
$\Delta \beta_{j}$, ($j$ = 1, 2, ...) have been propagated to the 
scattering 
amplitude,
and then to the differential cross section, providing

\begin{eqnarray}
\frac{d\sigma}{dq^2} \pm \Delta \left(\frac{d\sigma}{dq^2}\right).
\end{eqnarray}
By adding and subtracting the corresponding uncertainties we may estimate
the confidence region associated with all the extrapolations, which cannot 
be excluded
on statistical grounds. A typical result with ensembles I and II
is illustrated in Fig. 7, for $pp$ scattering at $\sqrt s$= 23.5 GeV. 
We see that,
as expected, the effect
of adding the experimental data at $\sqrt s$= 27.5 GeV (when statistically
justified) is to reduce drastically the uncertainty region. That result will
be fundamental in the extraction of the empirical information on
the eikonal, as shown in what follows.

\begin{figure}
\begin{center}
\includegraphics[width=8.0cm,height=7.0cm]{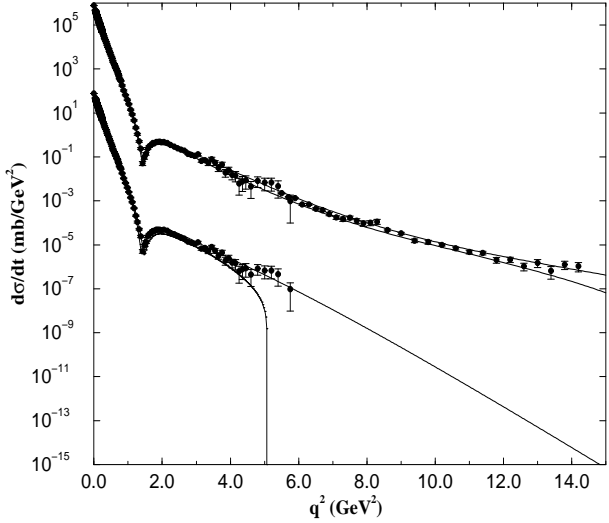}
\caption{Regions of uncertainties (limited by the
solid lines) in fits to $pp$ differential
cross section data, Eq. (23), at $\sqrt s = 23.5$ GeV with ensembles
I (below) and II (above) \cite{cmm03}.}
\end{center}
\end{figure}

- \leftline{\textit{ Eikonal in the momentum transfer space}}

By means of the Fourier transform, Eqs.(7-8), the parametrization
(21-22) provides analytical expressions for the real and
imaginary parts of the Profile function, $\Gamma_{R}(s,b)$ and
$\Gamma_{I}(s,b)$, and also the associated uncertainties.
From the fit results,
together with error propagation, we have found that

\begin{eqnarray}
\frac{\Gamma_{I}^{2}(s,b)}{[1-\Gamma_{R}(s,b)]^{2}} \ll1,
\nonumber
\end{eqnarray}
and therefore, the imaginary part of the
eikonal may be approximated by

\begin{eqnarray}
\chi_{I}(s,b)\approx\ln{1\over1-\Gamma_{R}(s,b)}
\end{eqnarray}
and the uncertainty $\Delta\chi_{I}$ determined directly from 
$\Delta\Gamma_{R}$
through propagation.

The next step is to go to the momentum transfer space
and concerns problem (2):
the Fourier transform can not be performed analytically
and therefore also the error propagation. For this reason
we used a semi-analytical method as follows.
Expanding the above equation, we express the remainder of the
series as

\begin{eqnarray}
R(s,b)=\ln[{1\over1-\Gamma_{R}(s,b)}]-\Gamma_{R}(s,b)
\end{eqnarray}
and then fit the numerical points (MINUIT) by a sum of Gaussians
in the impact parameter space:

\begin{eqnarray}
R_{\rm{fit}}(s,b)=\sum_{j=1}^{6} A_{j}e^{-B_{j}b^{2}}.
\end{eqnarray}
A typical result is displayed in Fig. 8.

\begin{figure}
\begin{center}
\includegraphics[width=8.0cm,height=7.0cm]{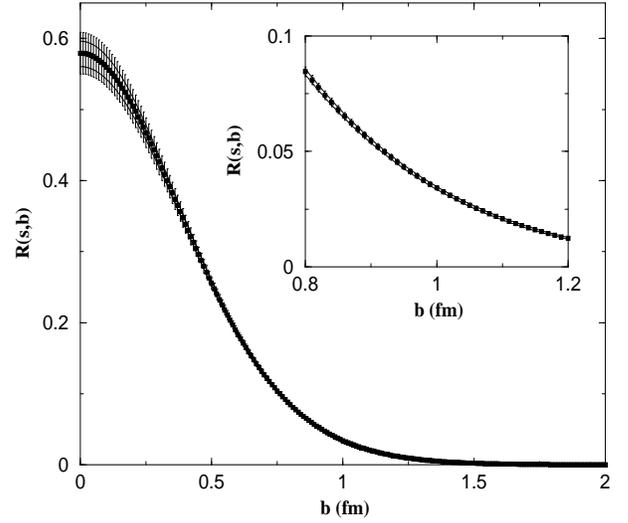}
\caption{Typical parametrization for the generated remainder $R(s,b)$ 
by means of Eq. (26) \cite{cmm03}.}
\end{center}
\end{figure}

With this, the errors
$\Delta A_{j}$ and $\Delta B_{j}$ may be propagated
determining $\Delta R_{fit}(s,q)$ and then
$\chi_{I}(s,q) \pm \Delta\chi_{I}(s,q)$. In order to
check the results and approximations, we performed also 
numerical integration through the NAG routine.

\leftline{- \textit{Results}}

One of the main results extracted from this analysis is the
statistical evidence of eikonal zeros in the momentum transfer space,
first presented in \cite{cm}.
In order to investigate the position of the zeros and, mainly, to
determine the uncertainties in its values, we  consider
the expected behavior of $\chi_{I}$ at large $q^2$, namely 
 $\chi_{I} \sim q^{-8}$.
In Fig. (9) we show a typical plot of the quantity $q^{8}\chi_{I}(s,q)$ as 
function of $q^2$. The shaded areas correspond to the uncertainties
obtained from error propagation.
This example shows clearly the role and the effect of data at large values of
the momentum transfer. In fact, within the uncertainties, ensemble I
does not allow to infer a zero,
but with ensemble II, we find statistical evidence for the change of sign.

\begin{figure}
\begin{center}
\includegraphics[width=6.0cm,height=5.0cm]{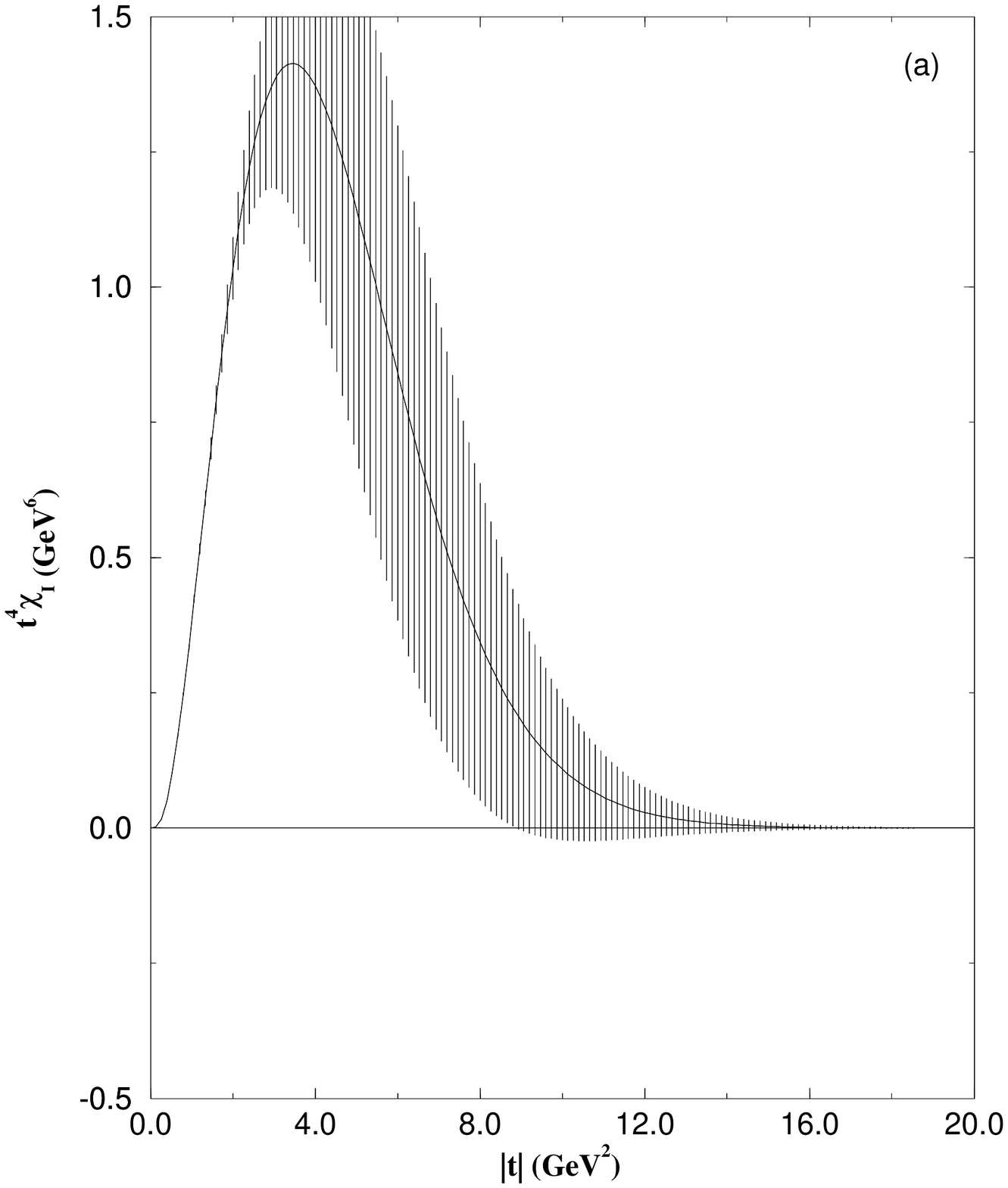}
\vspace{0.5cm}
\end{center}
\begin{center}
\includegraphics[width=6.0cm,height=5.0cm]{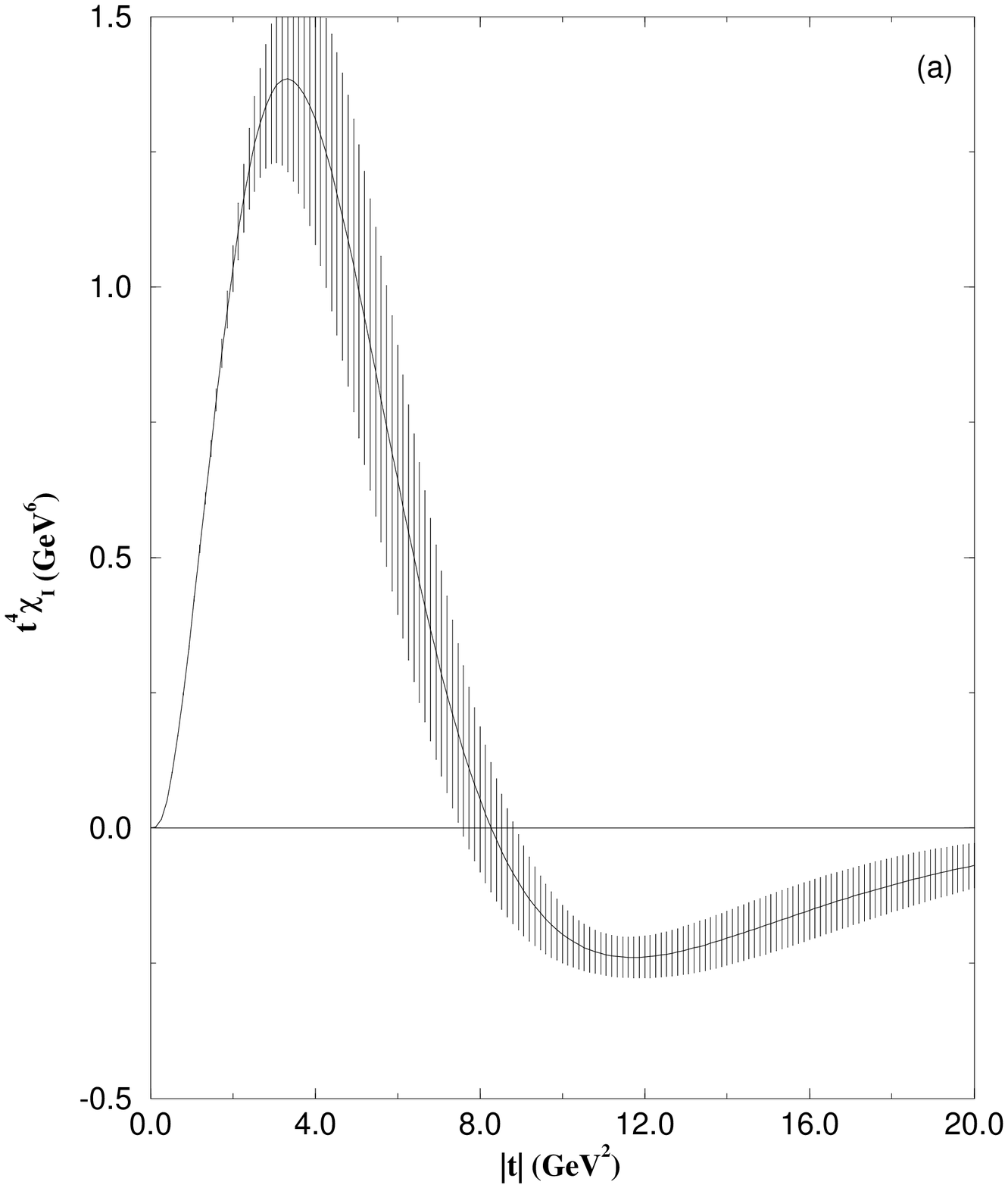}
\caption{The eikonal in the transfer momentum space (multiplied
by $t^4$) for $pp$ at $\sqrt s = 30.7$ GeV with ensemble I
(above) and II (below) \cite{cmm03}.}
\end{center}
\end{figure}

From plots like that we can determine
the positions of the zeros and the associated errors from the extrems
of the uncertainty region (in general not symmetrical). The position
of the zero can also be obtained from the numerical method,
but without uncertainties.

In Figure 10 it is shown the position of the zeros as function
of the energy determined by means of both the semi-analytical
(with uncertainties) and numerical (without uncertainties) methods.
Despite the systematic difference on the values with these methods,
we may conclude that the position of the zero decreases as the
energy increases. Roughly, $q^2_0:\ 8.5 \rightarrow 6.0$ GeV$^2$
as $\sqrt s :\ 20 \rightarrow 60$ GeV.

\begin{figure}
\begin{center}
\includegraphics[width=8.0cm,height=7.0cm]{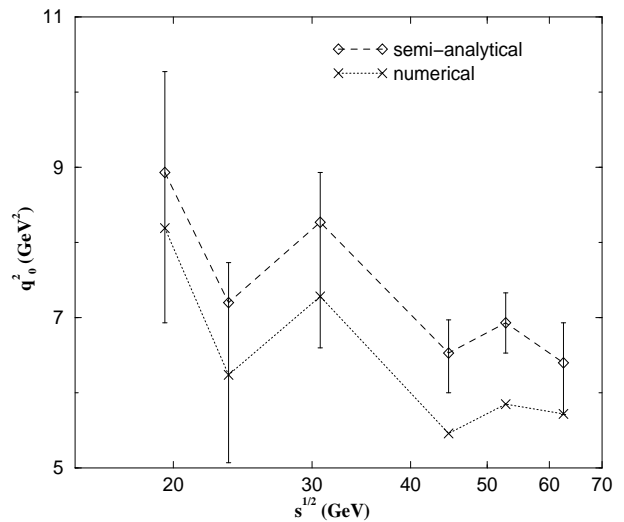}
\caption{Position of the eikonal zero in the momentum
transfer space as function of the energy \cite{cmm03}.}
\end{center}
\end{figure}

\leftline{- \textit{Discussion}}

As reviewed in \cite{cmm03}, there has been previous indication of eikonal
zeros in the momentum transfer space, but without associated
uncertainties. Our first statistical evidence, published in 1997,
indicated the position of the zero at $q^2_0 = 7 \pm 2$ GeV$^2$ 
\cite{cm}. In 2000, experiments performed at the Jefferson Laboratory, 
on electron-proton
scattering, have indicated an unexpected decrease of the ratio
between the electric and magnetic proton form factors
as the momentum transfer increases from 0.5 to
5.6 GeV$^2$. Moreover, extrapolations from empirical
fits indicate a change of sign (zero) in this ratio,
just at $q^2 \approx $ 7.7 GeV$^2$. Since for 
$pp$/$\bar{p}p$ scattering the eikonal is connected with the
hadronic matter form factor (see Sec. V, Eq. (38)), the above 
results on the position of the
zeros suggest novel and important insights on possible correlations 
between
hadronic and electromagnetic interactions. We discuss that subject in 
\cite{mmmhadron04}, calling attention to the possibility of
\textit{hadronic form factors depending on the energy}.

We have also obtained the value of the imaginary part of
the Eikonal at zero momentum transfer, that is, the central
opacity. The results are displayed in Fig. 11. As discussed in 
\cite{cmmm},
one naive way to test these results is with the Glauber
model for the scattering involving
hadrons $A$ and $B$, and the Optical Theorem at the elementary level.
In that case, the elementary cross section may be expressed
by

\begin{eqnarray}
\sigma_{elem}(s) = \frac{4\pi}{N_A N_B} \chi_{I}(s, q=0),
\nonumber 
\end{eqnarray}
where
$N_A$ and $N_B$ are the number of constituents in hadrons $A$ and $B$,
respectively.
If we take $N_A = N_B = 3$ we obtain $\sigma_{elem} \sim 6$ mb
at the ISR region, a result in agreement with other estimations.

\begin{figure}
\begin{center}
\includegraphics[width=8.0cm,height=7.0cm]{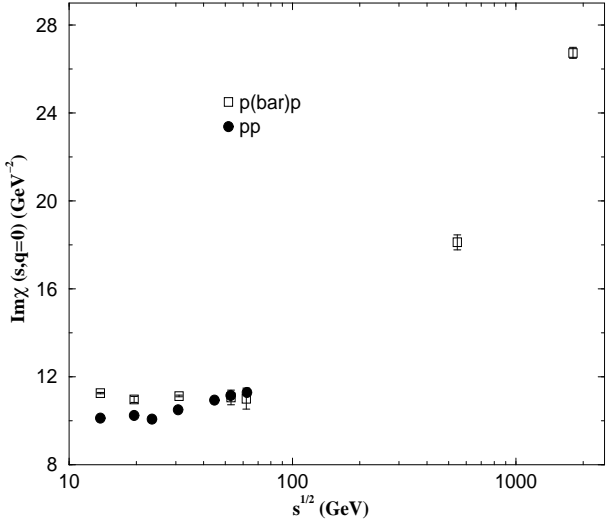}
\caption{Imaginary part of the eikonal at zero momentum transfer,
as function of the energy, from analysis on $pp$ and $\bar{p}p$
scattering \cite{cmm03}.}
\end{center}
\end{figure}

In Ref. \cite{cmm03} we discuss the applicability of our results in the 
phenomenological context, 
outlining some connections with nonperturbative QCD and
presenting a critical review on the
main results concerning ``model-independent" analyses.

\subsubsection{Constrained Fits and Energy Dependence}

Despite the results obtained with the parametrization discussed
in the last subsection, due to the fit procedure, we do not have the 
dependence on the energy of the free parameters $\alpha_i$, $\beta_i$.
Presently, we are investigating that subject and we review here some
preliminary results \cite{acmmhadron04}.

The energy dependence has been introduced according to some empirical
information: the increase of the total cross section
and of the slope parameter with $\ln^2 s$ and $\ln s$, respectively
(see Figs 1 and 3).
Let us consider the standard exponential parametrization 
for the imaginary part of the amplitude, normalized as

\begin{eqnarray}
\frac{\im F(s,q^2)}{s}={\sum_{i=1}^n}{\alpha}_i\exp[-\beta_iq^2].
\end{eqnarray}

At $q^2 = 0$, from the optical theorem, Eq. (18), we expect a
dependence of the parameters $\alpha_i$ with $\ln^2 s$, and the
slope represented by the parameters $\beta_i$ with $\ln s$.
These are the central choices in our approach. In order to treat
$pp$ and $\bar{p}p$ scatterings, in agreement with  Analyticity
and Crossing, we introduce crossing even and odd amplitudes and 
make use of the derivative dispersion relations,
Eqs. (15) and (16), to connect real
and imaginary parts of the amplitudes involved.

Specifically, for the imaginary part of the
scattering amplitude we consider the parametrizations

\begin{eqnarray}
\frac{\im F_{pp}(s,q^2)}{s}={\sum_{i=1}^n}{\alpha}_i(s)
\exp[-\beta_i(s)q^2],
\end{eqnarray}

\begin{eqnarray}
\frac{\im F_{\bar{p}p}(s,q^2)}{s}={\sum_{i=1}^n}{\bar{\alpha}}_i(s)
\exp[-\bar{\beta}_i(s)q^2],
\end{eqnarray}
and, based on the above arguments, we introduce the following general 
dependences
on the energy

\begin{eqnarray}
\left\{\begin{array}{l@{\quad\quad}}
\label{6}\alpha_i(s)=A_i+B_i\ln(s)+C_i\ln^2(s)\\
\beta_i(s)=D_i+E_i\ln(s)
\end{array}\right.
\end{eqnarray}
for $pp$ scattering and

\begin{eqnarray}
\left\{\begin{array}{l@{\quad\quad}}
\label{7}\bar{\alpha}_i(s)=\bar{A}_i+\bar{B}_i\ln(s)+\bar{C}_i\ln^2(s)\\
\bar{\beta}_i(s)=\bar{D}_i+\bar{E}_i\ln(s)
\end{array}\right.
\end{eqnarray}
for $\bar{p}p$ scattering, where $i=1,2,...n$. Defining the
crossing even (+) and odd (-) amplitudes,

\begin{eqnarray}
\im F_{+}(s,q^2)=\frac{\im F_{pp}(s,q^2)+
\im F_{\bar{p}p}(s,q^2)}{2},
\end{eqnarray}

\begin{eqnarray}
\im F_{-}(s,q^2)=\frac{\im F_{pp}(s,q^2)-
\im F_{\bar{p}p}(s,q^2)}{2}.
\end{eqnarray}
the corresponding real parts can be determined by means of the 
leading terms of the DDR, Eqs. (15-16), and so the corresponding real 
parts of the
$pp$ and $\bar{p}p$ amplitudes. With these analytic amplitudes we
obtain the differential cross section:

\begin{eqnarray}
\frac{d{\sigma}}{dq^2}=\frac{1}{16\pi s^2}
|\re F(s,q^2)+i\im F(s,q^2)|^{2}.
\end{eqnarray}

In order to treat simultaneous fits to $pp$ and $\bar{p}p$ data
we have considered only sets available at nearly the same energy,
namely $\sqrt s \sim$ 19.5, 31, 53 and 62 GeV. As a preliminary test
we make use of data at the diffraction peak, outside the Coulomb-nuclear 
interference region, 0.01 GeV$^2 < q^2 \leq $ 0.5 GeV$^2$, and 
the data providing the optical point, 

\begin{eqnarray}
\frac{d\sigma(s,q^2 = 0)}{dq^2} =
\frac{\sigma_{tot}(1+\rho^2)}{16\pi}.
\end{eqnarray}

We have performed simultaneous fits to the experimental data through the
MINUIT program. For this ensemble we used only two exponentials for
the imaginary part of the amplitude,
obtaining good reproduction of all the data
analyzed, as shown in Fig. 12. In Ref. \cite{acmmhadron04} we
also display the predictions for the differential cross sections at
the RHIC energies.
We are presently investigating the extension
of the analysis to the region of higher momentum transfer.

\begin{figure}
\begin{center}
\includegraphics[width=8.0cm,height=7.0cm]{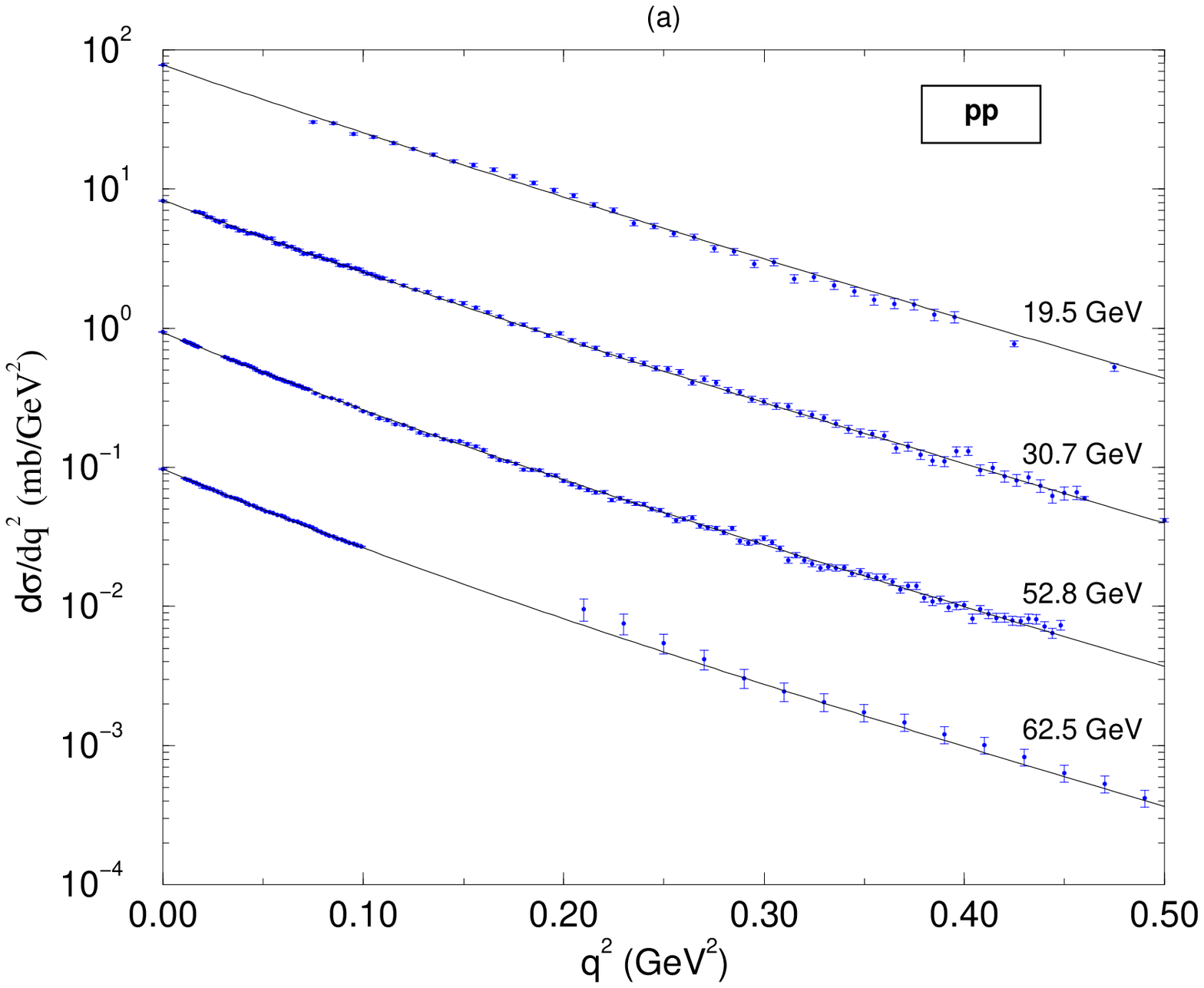}
\end{center}
\begin{center}
\includegraphics[width=8.0cm,height=7.0cm]{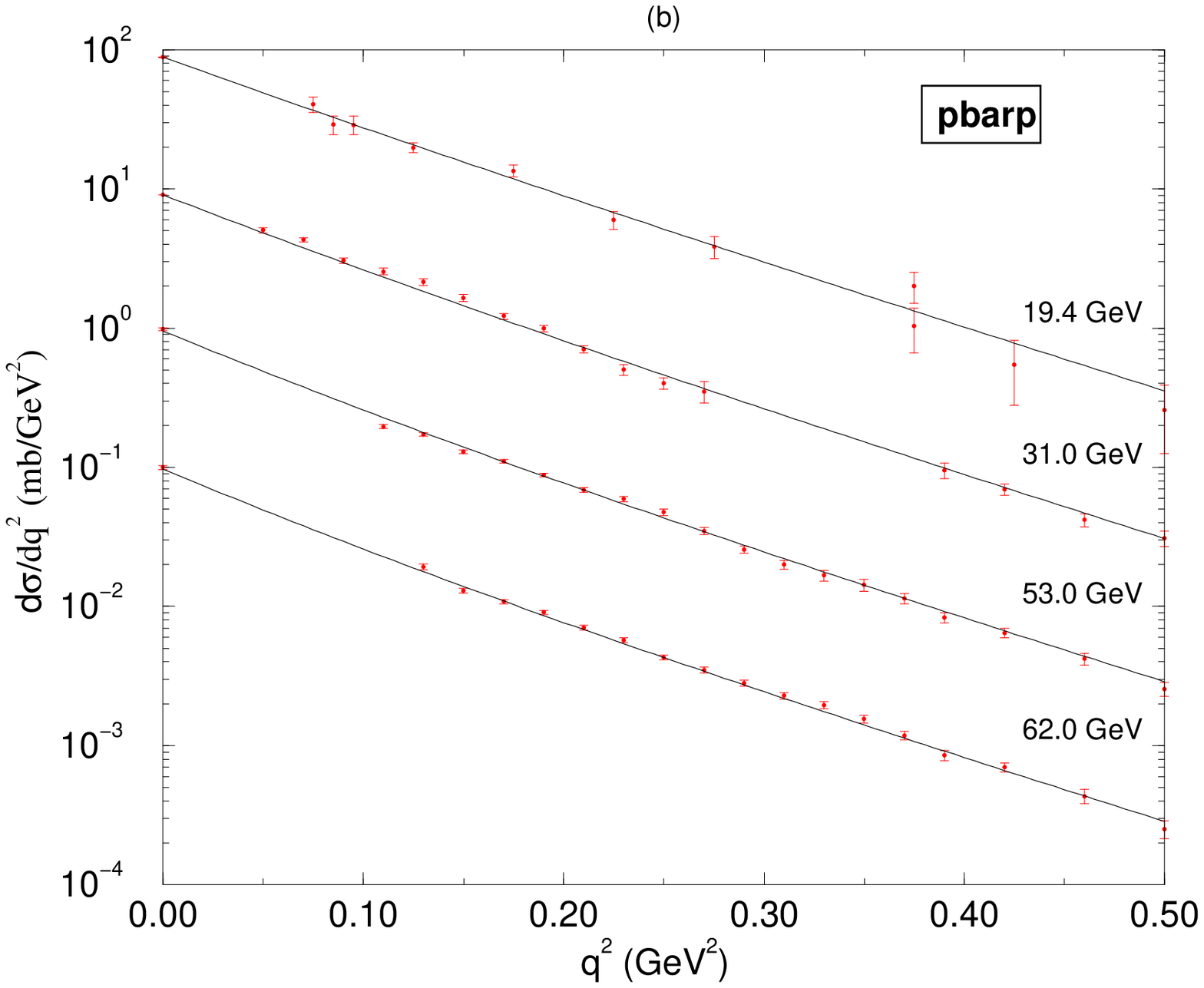}
\caption{Differential cross section at the diffraction peak: fit results
  and experimental data (displaced by a factor of 10) \cite{acmmhadron04}.}
\end{center}
\end{figure}

\subsection{Total Cross Sections and Slopes}

Another important quantity that characterizes the elastic hadron-hadron
scattering is the slope parameter, defined in Eq. (4). In practice it
may be determined by means of fits to the hadronic differential cross
section data in the region of small momentum
transfer, with the parametrization

\begin{eqnarray}
\frac{d\sigma}{dt} = \left[\frac{d\sigma}{dt}\right]_{t=0}
e^{-B|t|},
\end{eqnarray}
and, in general, it is connected with $\rho$ and $\sigma_{tot}$ through fits
in the region of Coulomb interference \cite{bc}.
The slope and the total cross section are also important quantities
in the determination of $\sigma_{tot}^{pp}$ from $\sigma^{p-air}$
(cosmic-ray experiments) but, as commented before, the procedure is
strongly model dependent. One reason is associated with the use of the
Glauber multiple diffraction formalism, in which $\sigma_{tot}(s)$
and $B(s)$ take part in the parametrization of the elastic amplitude,

\begin{eqnarray}
F^{pp}(s,t) \propto \sigma_{tot}^{pp}(s) \exp 
\left\{ \frac{B(s) t}{2}\right\}. 
\end{eqnarray}

As commented in \cite{alm03} and \cite{mmmbjp04}, different models
predict different relations between $\sigma_{tot}(s)$
and $B(s)$ and that is mirrored in the final value of the cross section,
contributing to the discrepancies already discussed.

Based on the above observations, we have investigated the
possibility to extract an empirical correlation between the
experimental data on $\sigma_{tot}(s)$
and $B(s)$, from $pp$ and $\bar{p}p$ scatterings. 

For the slope parameter, we have selected the data above 
the region of Coulomb-nuclear
interference and below
the ``break" in the hadronic slope at the diffraction
peak (localized at
$|t| \sim$ 0.2 GeV$^2$ at the ISR and Collider energies),
namely 0.01 $ < |t| < $ 0.20 GeV$^2$ (Fig. (3)). In this region, the
differential cross section data are well fitted by a single
exponential and therefore there is no change in the slope
associated with the $t$-dependence. For each energy we have compiled the
corresponding data on the total cross section.

Once more, the choice for a parametrization was based on the empirical 
observation that at high energies $B(s)$ increases with the logarithm
of $s$. Since the Kang-Nicolescu parametrization for the
total cross section is expressed in terms of $\ln s$ (Sec. III.B.2), 
we replaced this dependence by the slope parameter:

\begin{eqnarray}
\sigma_{tot}^{pp}(s) &=& c_1 + c_2B + c_3B^2,
\nonumber \\
\sigma_{tot}^{\overline{p}p}(s) &=& c_1 + c_2B +c_3B^2 +
c_4 e^{-B/2},
\label{KNeqs}
\end{eqnarray}
where $c_i$, $i = 1,2,3,4$ are free fit parameters.
That is a strictly mathematical choice, having 
nothing to do with the physics or model concept behind the  
Kang-Nicolescu parametrization.

Fits to the experimental data
have been performed with the 
CERN-MINUIT program and the results are displayed in Fig. 13.
It is expected that extrapolations to cosmic-ray energies
may be useful in the determination of the $pp$ total cross
section from $p$-air cross section, allowing to connect
$\sigma_{tot} - B$ in an almost model independent way. We are
presently investigating this subject.

\begin{figure}
\begin{center}
\includegraphics[width=8.0cm,height=7.0cm]{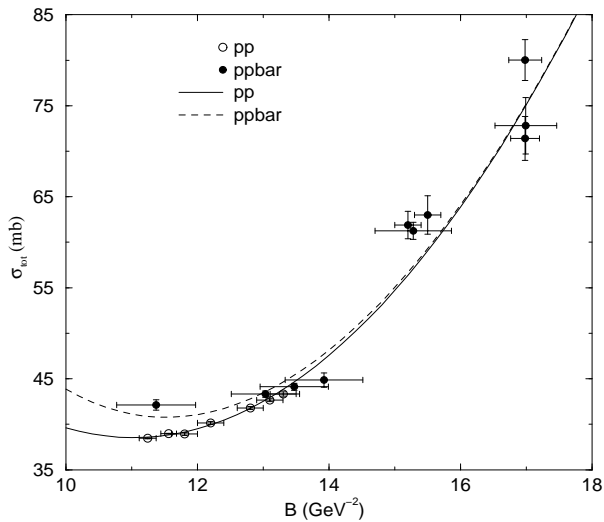}
\caption{Total cross sections in terms of the slope, and
the parametrization (38) \cite{mmmbjp04}.}
\end{center}
\end{figure}

In Ref. \cite{mmmbjp04} we also made use of the Donnachie-Landshoff
parametrization, which predicts a faster increase of the total cross
section as function of the slope parameter. Moreover, in \cite{mmmbjp04}
we also present a critical discussion on the recent measurement of
the slope parameter at the BNL RHIC, at 200 GeV, by the pp2pp Collaboration.
We call attention to the fact that
the combination $B =$ 16.3 $\pm$ 1.8 GeV$^{-2}$
and $\sigma_{tot} =$ 51.6 mb, indicated by the pp2pp analysis,
is in disagreement with the general trend for the
behaviors of $\sigma_{tot}$ and $B$. If this ``peer" is
correct, new physics is necessary. 
Using the above
$B$ value as input in our parametrizations, the corresponding
values of the total cross sections show agreement with
the $\sigma_{tot}$ versus $B$ data. However, these inferred
values for $\sigma_{tot}$ indicate new physics when plotted
as function of the energy. We conclude that
if this
measurement is correct and represents an hadronic quantity,
its high value  may indicate a ``break" in the slope near
0.02 GeV$^2$, a phenomenon that was never observed in both
$pp$ and $\bar{p}p$ scattering,
at $\sqrt s \leq $ 62.5 GeV and $\sqrt s \leq $ 1.8 TeV,
respectively and therefore, once more,
new physics is necessary.

\section{Eikonal Models}

It is expected that the eikonal function in the
momentum transfer space, $\chi(s,q)$, may be connected with some 
microscopic 
aspects of the underlying field theory (elementary interactions,
form factors, structure functions) and,  as
mentioned before, it corresponds to a unitarized scheme connected
with the experimental data.
Eikonal models are characterized by different choices for 
$\chi(s,q)$. In what follows we discuss our
results and researches through two eikonal models 
(geometrical and QCD-based)

\subsection{Geometrical Model - Inelastic Channel}

In this subsection we review the description of $\bar{p}p$ 
multiplicities
distributions (inelastic channel) from models for the elastic channel
in the context of the geometrical picture (contact interactions).

\leftline{- \textit{ Elastic and Inelastic Channels}}

Through the Unitarity and the Inelastic Overlap Function,
defined in Sec. II.C, we can connect elastic and
inelastic scattering. This is done by expressing 
the topological cross section
for producting an even number $n$ of charged particles
at $s$ in terms of $G_{in}$:

\begin{eqnarray}
\sigma_{n}(s)= \int d^{2}{\bf b}\ \sigma_{n}(s,b) = 
\int d^{2}{\bf b}\ 
G_{in}(s,b) \left[ \frac{\sigma_{n}(s,b)}{G_{in}(s,b)} \right].
\nonumber
\end{eqnarray}

If $n(s)$  and $<n>_{(s)}$ are the hadronic and averaged multiplicities,
respectively, by introducting the KNO variable 
$Z=n(s)/<n>_{(s)}$,
the hadronic multiplicity distribution may be expressed by

\begin{eqnarray}
\Phi (s,Z)= <n>_{(s)} \frac{\sigma_{n}(s)}{\sigma_{in}(s)} 
= \frac{\int d^{2} {\bf b} \frac{G_{in}(b,s)}{r(b,s)} 
\varphi 
(\frac{Z}{r(b,s)})}{ \int d^{2} {\bf b}\ G_{in} (b,s)},
\nonumber
\end{eqnarray}
where $\varphi$ is the elementary multiplicity distribution
and $r(b,s) = <n>_{(b,s)}/<n>_{(s)}$ the elementary multiplicity
function. 

In Ref. \cite{bmv}, in the context of the
geometrical picture, the elementary contact interaction
process was based on $e^+e^-$ scattering data. In this
approach we express 

\begin{eqnarray}
r(s,b)=\xi(s) \chi_{I}^{\gamma}(s,b), \nonumber
\end{eqnarray}
where 

\begin{eqnarray}
\xi(s)= \frac{\int d^{2} b\ G_{in}(s,b)}{\int d^{2} b\ 
G_{in}(s,b) \chi_{I}^{\gamma}(s,b)} \nonumber
\end{eqnarray}
and the power $\gamma$
is determined by fitting the average multiplicity from
 $e^+e^-$ scattering data through a power law parametrization: 

\begin{eqnarray}
<n>_{e^+ e^-} = 
A[\sqrt s]^{\gamma}. \nonumber
\end{eqnarray}

The elementary distribution
$\varphi(Z/r(b,s))$ is represented by a Gamma distribution
and determined also by fits to $e^+e^-$ data.

With inputs for $G_{in}(s,b)$ and/or $\chi_{I}(s,b)$, obtained from fits
to elastic scattering data,
we have no free parameter and
the hadronic multiplicity distribution as function
of $Z$ and $s$ may be inferred.

\leftline{- \textit{ Elastic-channel inputs and results}}

In Ref. \cite{bmv} we made use of three inputs from the elastic
sector. Two are based on the Multiple Diffraction Formalism,
in which the eikonal in the momentum transfer space is expressed by

\begin{eqnarray}
\chi (b,s)= C\int qdqJ_{0}(qb)G_{A}G_{B}f, 
\end{eqnarray}
where $G_{A}$ and  $G_{B}$ are the hadronic form
factors, $f$ the elementary (constituent - constituent) amplitude
and $C$ does not depend on the transferred momentum. In this case
we made use of the parametrizations used by  Chou and Yang
and also by Menon and Pimentel. Both present good descriptions
of the experimental data in the elastic channel. The other input
corresponds to the Short Range Expansion of the inelastic overlap 
function, 
introduced by Henzi and Valin, 

\begin{eqnarray}
G_{in}(b,s)=P(s)\exp \{-b^{2}/4B(s)\}k(x,s), \nonumber
\end{eqnarray} 
with $k$ being expanded in terms of a short-range variable $x=b\ 
\exp \{-(\epsilon b)^{2}/4 B(s)\}$, i.e.

\begin{eqnarray}
k(x,s)=\sum_{n=0}^{N} \delta_{2n}(s) \left[ \frac{\epsilon\ 
\exp \{1/2\}}{\sqrt{2B(s)}}\ x \right]^{2n}. \nonumber
\end{eqnarray}

With particular parametrizations excellent agreement with experimental 
data on $pp$ and 
$\overline{p}p$ elastic scattering is achieved, allowing to infer the
black-edge-large (BEL) behavior.

In Ref. \cite{bmv} a detailed discussion is presented on several
variants from the elastic channel and parametrizations from
$e^-e^+$ scattering. In particular, the results for the
multiplicities distributions, with the BEL inelastic overlap
function, at
$\sqrt s =$ 52.6 and 546 GeV are displayed in Fig. 14,
together with the experimental data. 
The prediction shows that
the violation of the KNO scaling is well described.

\begin{figure}
\begin{center}
\includegraphics[width=8.0cm,height=7.0cm]{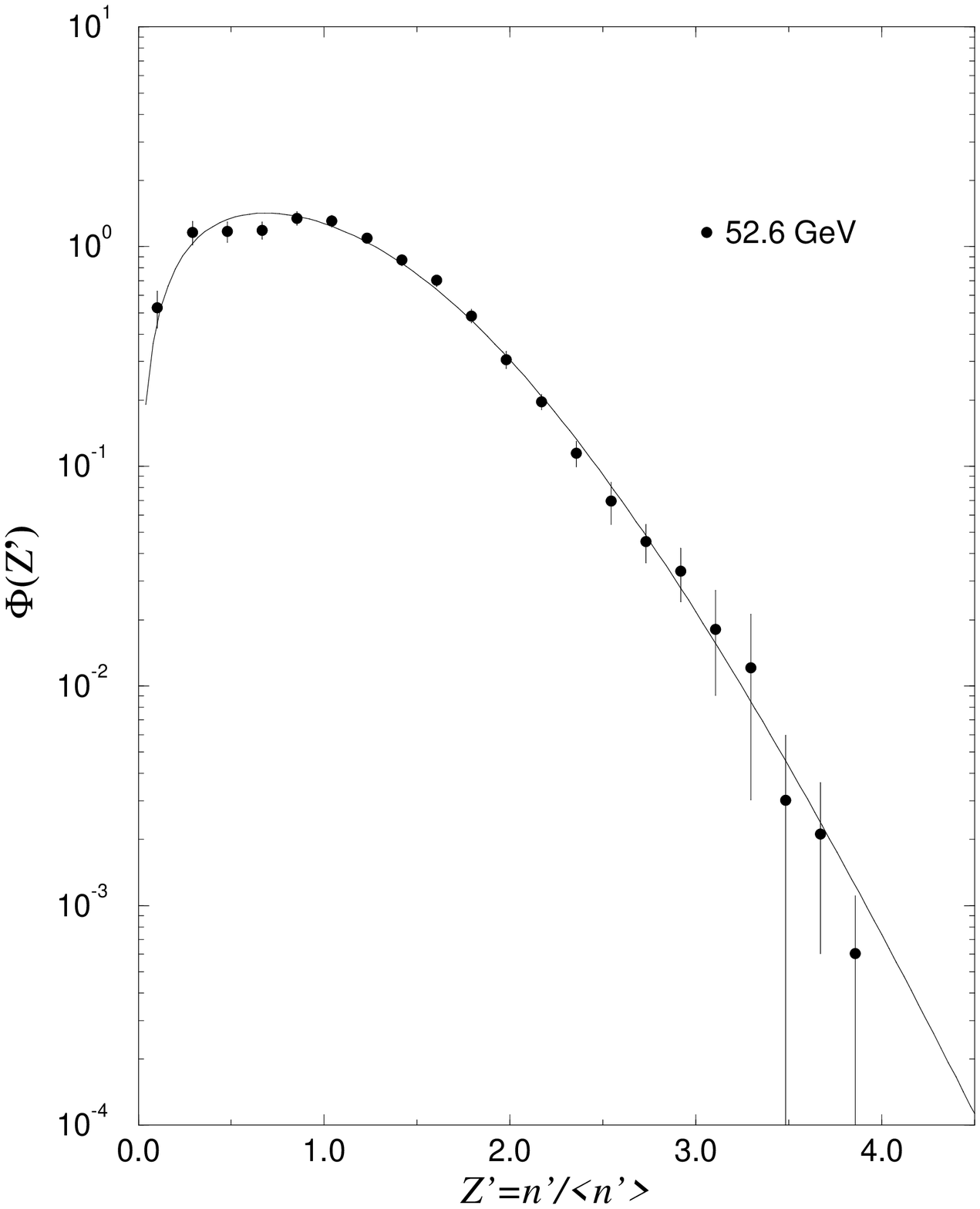}
\end{center}
\vspace{0.3cm}
\begin{center}
\includegraphics[width=8.0cm,height=7.0cm]{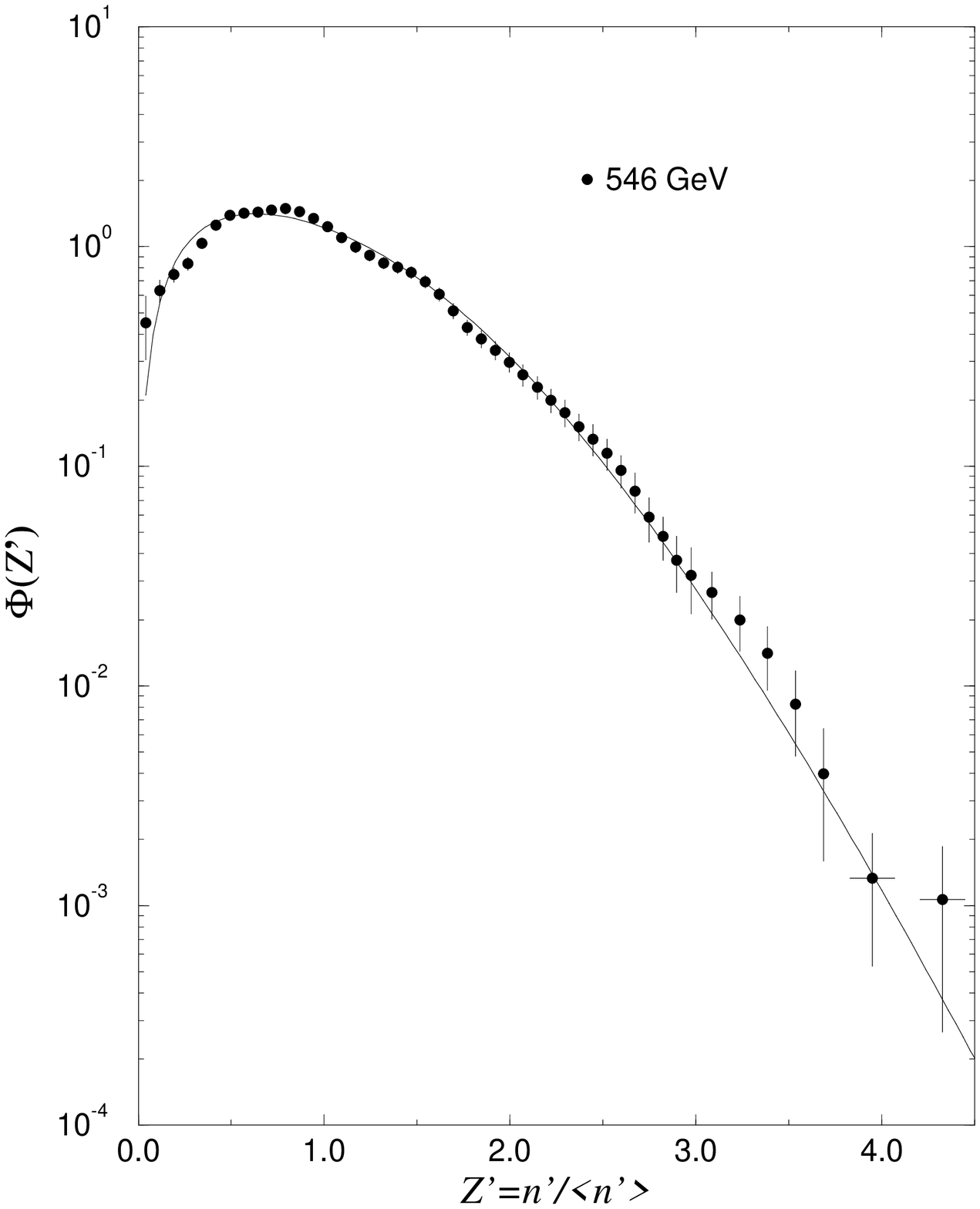}
\vspace{0.3cm}
\caption{Scaled multiplicity distribution for inelastic $pp$ data at
$\sqrt s =$ 52.6 GeV and $\bar{p}p$ data at 546 GeV, compared with the
model predictions \cite{bmv}.}
\end{center}
\end{figure}

\subsection{QCD-inspired models}

In this section we outline some research in course with
the eikonal approach in connection with some QCD concepts.
As we shall discuss the main point concerns the
gluon-gluon contributions to the hadronic cross sections
which we have investigated either from a dynamical gluon mass
approach or by introducting the momentum scale in the gluonic distribution 
functions.
After a review on the basic formalism we outline some aspects
of both approaches.

\subsubsection{Basic formalism}

The formalism was introduced by Afek, Leroy, Margolis and
Valin \cite{qcdb1} and developed by several authors, including (for
our purposes) Durand, Pi \cite{dp}, Block, Gregores, Halzen and Pancheri 
\cite{qcdb2}.

Originally, the point was to separate contributions from
soft (S) and semi-hard (SH) inelastic processes by expressing

\begin{eqnarray}
G_{inel}(s,b) &=& 1 - \bar{P}_{S} \bar{P}_{SH} \nonumber \\
&=&
1 - e^{- 2Re\ \chi_{S}(s,b)} e^{- 2Re\ \chi_{SH}(s,b)},
\nonumber
\end{eqnarray}
where $\bar{P}_{S}$ is the probability of $NO$ soft inelastic 
process and $\bar{P}_{SH}$ the probability of $NO$ semi-hard 
inelastic process. 
Therefore, that indicated an additive contribution in the eikonal:
$\chi(s,b) = \chi_{S}(s,b) + \chi_{SH}(s,b)$.

In the recent version by Block et al. \cite{qcdb2}
different elementary contributions from quarks and gluons 
have been introduced:
the gluon-gluon contribution comes
from the parton model, the
quark-quark from regge parametrization and the quark-gluon 
by phenomenological inputs. In what follows we shortly review the
main formulas.

The normalization for the eikonal reads

\begin{eqnarray}
F(s, q) = ik \int_{0}^{\infty} bdb J_{0}(qb)
\left[ 1 - e^{i\chi(s,b)} \right], \nonumber 
\end{eqnarray}

\begin{eqnarray}
\chi(s,b) = \re \chi(s,b) + i \im \chi(s,b). \nonumber
\end{eqnarray}

For $pp$ and $\bar{p}p$ scattering the crossing even and odd 
contributions are 
expressed by
$\chi_{p\bar{p}} = \chi^+ + \chi^-$ and 
$\chi_{pp} = \chi^+ - \chi^-$. The odd eikonal is assumed not 
to contribute at 
the asymptotic energies
and is parametrized by

\begin{eqnarray}
\chi^{-}(s,b) = C^{-} \frac{m_0}{\sqrt s}e^{i\pi/4}w(b,\mu_{odd}). 
\nonumber
\end{eqnarray}

Analyticity (generation of real and imaginary parts) for the
even part is assumed as given by the prescription

\begin{eqnarray}
\chi^{+}(s,b) \quad &\Rightarrow& \quad
\chi^{+}(se^{-i\pi/2},b) = \nonumber \\ 
&=& Re\ \chi^{+}(s,b) + i Im\ \chi^{+}(s,b).
\nonumber 
\end{eqnarray}

The even eikonal is expressed as a sum of three contributions, from
quark-quark ($qq$), quark-gluon ($qg$) and
gluon-gluon ($gg$) interactions,

\begin{eqnarray}
\chi^+(s,b) = \chi_{qq}(s,b) + \chi_{qg}(s,b) + \chi_{gg}(s,b), \nonumber
\end{eqnarray}
which individually factorize in $s$ and $b$,

\begin{eqnarray}
\chi_{ij}(s,b) = i \sigma_{ij}(s) w(b, \mu_{ij}), \nonumber
\end{eqnarray}
where $i,j = q,g$.

The {\it impact parameter distribution function} for each process comes from
convolution involving dipole form factors (Chou-Yang Model):

\begin{eqnarray}
w_{ii}(b, \mu_{ii})  = \int d^2\vec{b}' \rho_{i}(|\vec{b}'|
\rho_{i}(|\vec{b} - \vec{b}'|), \nonumber 
\end{eqnarray}

\begin{eqnarray} 
\rho(b) =  <G(q)> = < \frac{1}{(1 + q^2/\mu^2)^2} >, \nonumber
\end{eqnarray}
where the angular brackets denote the symmetrical two-dimensional 
Fourier transform. Therefore,

\begin{eqnarray}
w_{ii}(b) = \frac{1}{8} \frac{\mu_{ii}^2}{12\pi}[\mu_{ii} b]^3 
K_3(\mu_{ii} b),
 \nonumber 
\end{eqnarray}
and for $i \not= j$ it is assumed that

\begin{eqnarray}
\mu_{ij} = \sqrt{\mu_{ii} \mu_{jj}}.  \nonumber
\end{eqnarray}

The {\it elementary cross sections} for each process are introduced as
follows. The {\it quark-quark} contribution is parametrized as a 
constant plus a 
Regge
(even) term,

\begin{eqnarray}
\sigma_{qq}(s) = C + C_{R}^{+} \frac{m_0}{\sqrt s} \nonumber
\end{eqnarray}
and the {\it quark-gluon} term as

\begin{eqnarray}
\sigma_{qg}(s) = C_{qg} \log \frac{s}{s_0}. \nonumber
\end{eqnarray}

The {\it gluon-gluon} contribution is considered as the responsible for
the increase of the total cross section at the highest energies
and is
calculated through the parton model
approach,

\begin{eqnarray}
\sigma_{gg}(s) = c_{gg}\int_{0}^{1} d\tau F_{gg}(\tau) \hat\sigma_{gg}(\hat{s}), 
\end{eqnarray}
with
\begin{eqnarray}
F_{gg} = \int_0^1 \int_0^1 dx_1 dx_2 f_g(x_1) f_g(x_2) \delta(\tau - x_1x_2), 
\end{eqnarray}
where $f_g(x_i)$ is the gluon distribution function, $\tau = \hat{s}/s$, 
and the symbol
$\hat{ }$
denotes the elementary process. In \cite{qcdb2} the elementary cross
section is given by

\begin{eqnarray}
\hat\sigma_{gg}(\hat{s}) = 
\frac{9\pi \alpha_{s}^2}{m_{0}^2} \theta(\hat{s} - m_{0}^{2}),
\end{eqnarray}
implying a cutoff $m_0$ for the particle production threshold
and it is assumed the following simple parametrization for the gluon 
distribution function

\begin{eqnarray}
f_g (x) &=& N_g \frac{(1 - x)^5}{x^{1 + \epsilon}}, \nonumber \\
N_g &=& \frac{1}{2} 
\frac{(6 - \epsilon)(5 - \epsilon)...(1 - \epsilon)}{5!}. 
\end{eqnarray}

The model has 6 fixed parameters, $m_0$, $\epsilon$, $\mu_{qq}$,
$\mu_{gg}$, $\mu_{odd}$, $\alpha_s$ and 6 free parameters, determined 
from fits
to $pp$ and $p\bar{p}$ {\it forward} scattering data, namely
$\sigma_{tot}(s)$, $\rho(s)$ and $B(s)$ above 15 GeV \cite{qcdb2}.
We have shown that the model
applies only to forward and small momentum
transfer regions \cite{lmmhadron02}.

In the next two sections we shall discuss two researches in course
concerning the determination of the contribution from gluon-gluon
interactions, Eqs. (39-42).

\subsubsection{Dynamical gluon mass}

The possibility that the gluon propagator may be regularized
by a dynamically generated gluon mass \cite{cornwall} has
recently provided important phenomenological description
of several processes \cite{natale}. The approach allows to
calculate  the contribution for the elementary $gg$ cross section
Eq. (42) and the main point is the association of the mass scale
with the dynamical gluon mass.
The basic ingredients are the expressions for the
dynamical gluon mass,
 
\begin{eqnarray}
M^2_g(\hat{s}) =m_g^2 \left[
\frac{ \ln [(\hat{s}+4{m_g}^2) / \Lambda_{\mathrm{QCD}} ^2]}{ \ln 
[(4{m_g}^2) / \Lambda_{\mathrm{QCD}}^2] }\right]^{- 12/11},
\nonumber
\end{eqnarray}
and the associated running coupling constant

\begin{eqnarray}
\alpha_{s} (\hat{s})= \frac{4\pi}{\beta_0 \ln\left[
(\hat{s}+4M_g^2(\hat{s}))/\Lambda_{\mathrm{QCD}}^2 \right]},
\nonumber 
\end{eqnarray}
where $\beta_0 = 11- \frac{2}{3}n_f$ and $n_f$ is the number of 
flavors. Preliminary tests with these contributions, in the context
of the model described in the last subsection, have shown that the
experimental data on $\sigma_{tot}(s)$, $\rho(s)$ and $B(s)$ are well
described \cite{lmmmn}, as exemplified in Fig. (15). We are presently
investigating the contributions from the other elementary processes,
$qq$ and $qg$.

\begin{figure}
\begin{center}
\includegraphics[width=8.0cm,height=7.0cm]{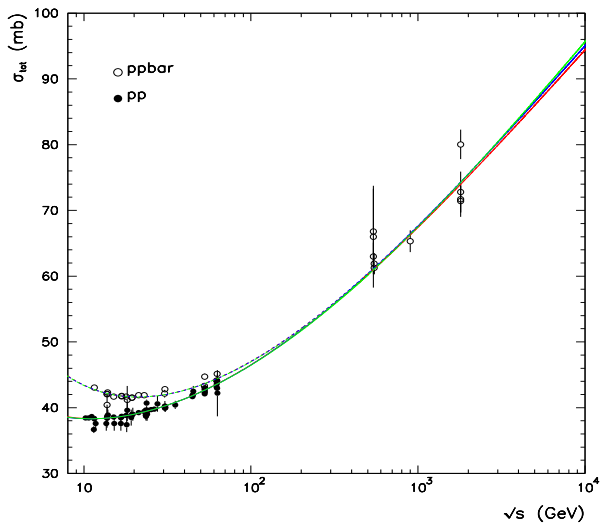}
\caption{Description of the total cross sections through the QCD-based
model with dynamical gluon mass $m_0 =$ 500, 600 and 700 MeV \cite{lmmmn}.}
\end{center}
\end{figure}

\subsubsection{Momentum  Scale}

Presently, we are attempting to improve
the descriptions of the QCD-inspired models by taking into account 
the momentum transfer scale
in the gluon distribution functions. The point is to replace
the simple choice in Eq. (42), by distribution functions with
the $Q^2$ dependence, namely

\begin{eqnarray}
f_{g}(x_i) \rightarrow f_{g}(x_i, Q^2). \nonumber
\end{eqnarray}

That can be implemented, following the approach by Durand and Pi \cite{dp},
by introducing the differential cross section

\begin{eqnarray}
\hat\sigma_{gg}(\hat{s}) = 
\int d|\hat{t}| \frac{d\hat\sigma_{gg}}{d|\hat{t}|}(\hat{s}, \hat{t}),
\nonumber
\end{eqnarray}
with
\begin{eqnarray}
\frac{d\hat\sigma_{gg}}{d|\hat{t}|}(\hat{s}, \hat{t}) &=&
\frac{9\pi\alpha_{s}^2}{2}
[\frac{3}{\hat{s}^2} + \frac{\hat{t}}{\hat{s}^3}
+ \frac{\hat{t}^2}{\hat{s}^4} + \frac{1}{\hat{t}^2} +
\frac{1}{\hat{s}\hat{t}} \nonumber \\ 
&-& \frac{\hat{t}}{\hat{s}(\hat{s} + \hat{t})^2}]
\nonumber
\end{eqnarray}
and by considering  $Q^2 = |\hat{t}|$.

The novel input concerns the updated determinations of the gluon distribution 
functions (CETEQ6), parametrized by means of
Chebyshev polynomials. Presently, the implementation in the QCD-inspired 
approach is being
developed.

\section{Nonperturbative QCD}

As commented before the difficulties associated with high-energy
soft processes arise 
from the fact that perturbative QCD can not be applied 
and presently we do 
not know  how to calculate even the elastic hadron-hadron scattering 
amplitudes from a pure nonperturbative 
QCD formalism. However, progresses have been achieved through the approach 
introduced by Landshoff and Nachtmann \cite{landotto}, developed by 
Nachtmann \cite{otto1} and connected with the
Stochastic Vacuum Model (SVM) (introductory reviews may be
found in \cite{rev}).
In particular, through this
formalism and in some restricted kinematic conditions, it is possible to 
connect the gluon two-point correlation function with elementary
(quark-quark) scattering amplitude. 

In this section we review the
results we have obtained for these amplitudes with correlators determined 
from lattice QCD and also in the context of the Constrained Instantons.
  
\subsection{Stochastic Vacuum Model}

The approach has its origins in the attempts by Landshoff and
Nachtmann to connect soft high-energy processes with nonperturbative
properties of the QCD vacuum, as for example, the {\it gluon condensate}
introduced by Shifman, Vainstein and Zakharov \cite{svz}. In the first
version \cite{landotto} quarks couple with Abelian gluons. The non-Abelian
version was developed by Nachtmann in the context of QCD and using the
eikonal method for high energy interactions \cite{otto1}.
The scattering amplitude is calculated by means of a functional integral
approach and is connected to a correlation function of two lightlike
Wegner-Wilson loops. These correlation functions can be evaluated through the
Stochastic Vacuum Model, in which the low frequency contributions 
to the functional integral of QCD are described in terms of a stochastic
process by means of a cluster expansion \cite{svm}. The model incorporates 
the gluon
condensate concept and assumes that the correlation of two field strengths
decreases rapidly with distance; due to an effective chromomagnetic monopole
condensate, the QCD vacuum acts as a dual superconductor.

In this formalism the low frequencies contributions in the functional 
integral of QCD are described in terms of a stochastic process, by means of 
a cluster expansion. The most general form 
of the lowest cluster is the gauge invariant two-point field 
strength correlator \cite{svm,dfk}

\begin{eqnarray}
&<&{\bf{F}}_{\mu \nu}^{\rm C}(x){\bf{F}}_{\rho
\sigma}^{\rm D}(y)>=  \nonumber \\ 
&=&{\delta}^{\rm CD}g^{2}\frac{<FF>}{12(N_c^2-1)}\{(
{\delta}_{\mu\rho}{\delta}_{\nu \sigma}-{\delta}_{\mu \sigma}
{\delta}_{\nu\rho}){\kappa}D({z^2/a}^{2})+ \nonumber \\
&+&
\frac{1}{2}[{\partial}_{\mu}(z_{\rho}{\delta}_{\nu
\sigma}-z_{\sigma}{\delta}_{\nu
\rho}) + \nonumber \\
&+&{\partial}_{\nu}(z_{\sigma}{\delta}_{\mu
\rho}-z_{\rho}{\delta}_{\mu \sigma})](1-{\kappa})D_{1}({z^2/a}^{2})\}, 
\nonumber
\end{eqnarray} 
where $z=x-y$ is the two-point distance, $a$ is a characteristic 
correlation length, ${\kappa}$ a constant,  $g^{2}<FF>$ the gluon 
condensate and $N_c$ the number of colours 
(${\rm C}, {\rm D}=1,...,N_c^2-1$). 
The two scalar functions $D$ and $D_{1}$ describe the correlations
and they play a central role in the application 
of the SVM to high energy scattering. 
Once one has information 
about $D$ and $D_{1}$, the SVM leads to the determination of the
 elementary 
quark-quark scattering amplitude, which constitutes important 
input for models
aimed to construct hadronic amplitudes. The main formulas are as
follows. 

The elementary amplitude $f$ in the momentum transfer space is expressed
in terms of the elementary profile $\gamma$
by

\begin{eqnarray}
f(q^2)=\int_0^{\infty}bdb J_0(qb)\gamma(b). 
\end{eqnarray}
In the Nachtmann approach the no-colour exchange 
parton-parton (loop-loop) amplitude can be written as 

\begin{eqnarray}
\gamma&=&{\langle}Tr[{\cal{P}}e^{-ig{\int}_{loop 1}d{\sigma}_{\mu
\nu}F_{\mu \nu}(x;w)}-1] \nonumber \\ 
& & Tr[{\cal{P}}e^{-ig{\int}_{loop2}d{\sigma}_{\rho\sigma}
F_{\rho \sigma}(y;w)}-1]{\rangle}, \nonumber
\end{eqnarray}
where ${\langle}{\rangle}$ means the functional integration over the 
gluon fields (the integrations are over the respective loop areas), 
and $w$ is a common reference point from which the integrations are 
performed. In the 
SVM by taking the Wilson loops on the light-cone the {\it leading order} 
contribution to the amplitude is given by 

\begin{eqnarray}
\gamma(b)=\eta{\epsilon}^{2}(b), 
\end{eqnarray}
where $\eta$ is a constant depending 
on normalizations  and

\begin{eqnarray}
\epsilon(b)=g^{2}{\int}{\int}d{\sigma}_{\mu
\nu}d{\sigma}_{\rho \sigma}Tr{\langle}F_{\mu \nu}(x;w)F_{\rho
\sigma}(y;w){\rangle}. \nonumber 
\end{eqnarray}

After a two-dimensional integration, $\epsilon(b)$ can be 
expressed in terms of the correlation functions 
by 

\begin{eqnarray}
\epsilon(b)=\epsilon_I(b)+\epsilon_{II}(b), 
\end{eqnarray}
where

\begin{eqnarray}
\epsilon_I(b)={\kappa}{\langle}g^2FF{\rangle}
{\int}_{b}^{\infty}db'(b'-b){\cal{F}}_{2}^{-1}
[D(k^2)](b'), 
\end{eqnarray}

\begin{eqnarray}
\epsilon_{II}(b)=({1-\kappa}){\langle}g^2FF{\rangle}
{\cal{F}}_{2}^{-1}[\frac{d}{dk^{2}}D_{1}(k^2)](b).
\end{eqnarray}
For 
${\cal{D}}=D$ or ${\cal{D}}=D_{1}$ we have
${\cal{D}}(k^2)={\cal{F}}_4[{\cal{D}}(z^2)]$, where
${\cal{F}}_n$ denotes a n-dimensional Fourier transform.

With the above formalism, once one has inputs for the correlation 
functions $D(z)$ and $D_1(z)$, the elementary amplitude in the 
momentum transfer space, Eq. (43),  may, in 
principle, be evaluated through Eqs. 
(44-47).
It is important to stress that, as constructed, the
formalism is intended for small momentum transfer ($q^2 \lesssim {\cal O}(1)$ 
GeV$^2$) and asymptotic energies $s \rightarrow \infty$.
Despite of these limitations, the investigation of soft high energy
scattering at the energies presently available  has led to satisfactory 
results
\cite{dfk,fp,berger}.

\subsection{Elementary Amplitudes}

In this section we review the results we have obtained from inputs
for the above correlators from lattice QCD
\cite{mmt,mm03}. We also comment the
research in course in the semi-classical context of Instantons
\cite{doro}.

\subsubsection{Correlators from Lattice QCD}

Numerical determinations of the above correlation 
functions, in limited interval of physical distances, exist from lattice 
QCD in both quenched approximation (absence of fermions)   
and full QCD (dynamical fermions included) \cite{digi}.

With the procedure described above (see \cite{mmt} for all the 
calculational details), 
the elementary scattering amplitude in the momentum transfer 
space can be 
determined in numerical form. 
In order to obtain analytical expressions, suitable for 
investigating distinct contributions and also for phenomenological uses, 
we have parametrized these numerical points through a sum of exponentials 
in $q^2$:

\begin{eqnarray}
\frac{f(q^2)}{f(0)}=\sum_{i=1}^n\alpha_i e^{-\beta_i q^2}.
\end{eqnarray}
The results are displayed in Fig. (16) from both quenched
approximation and full QCD, together with the corresponding exponential 
components.

\begin{figure}
\begin{center}
\includegraphics[width=8.0cm,height=7.0cm]{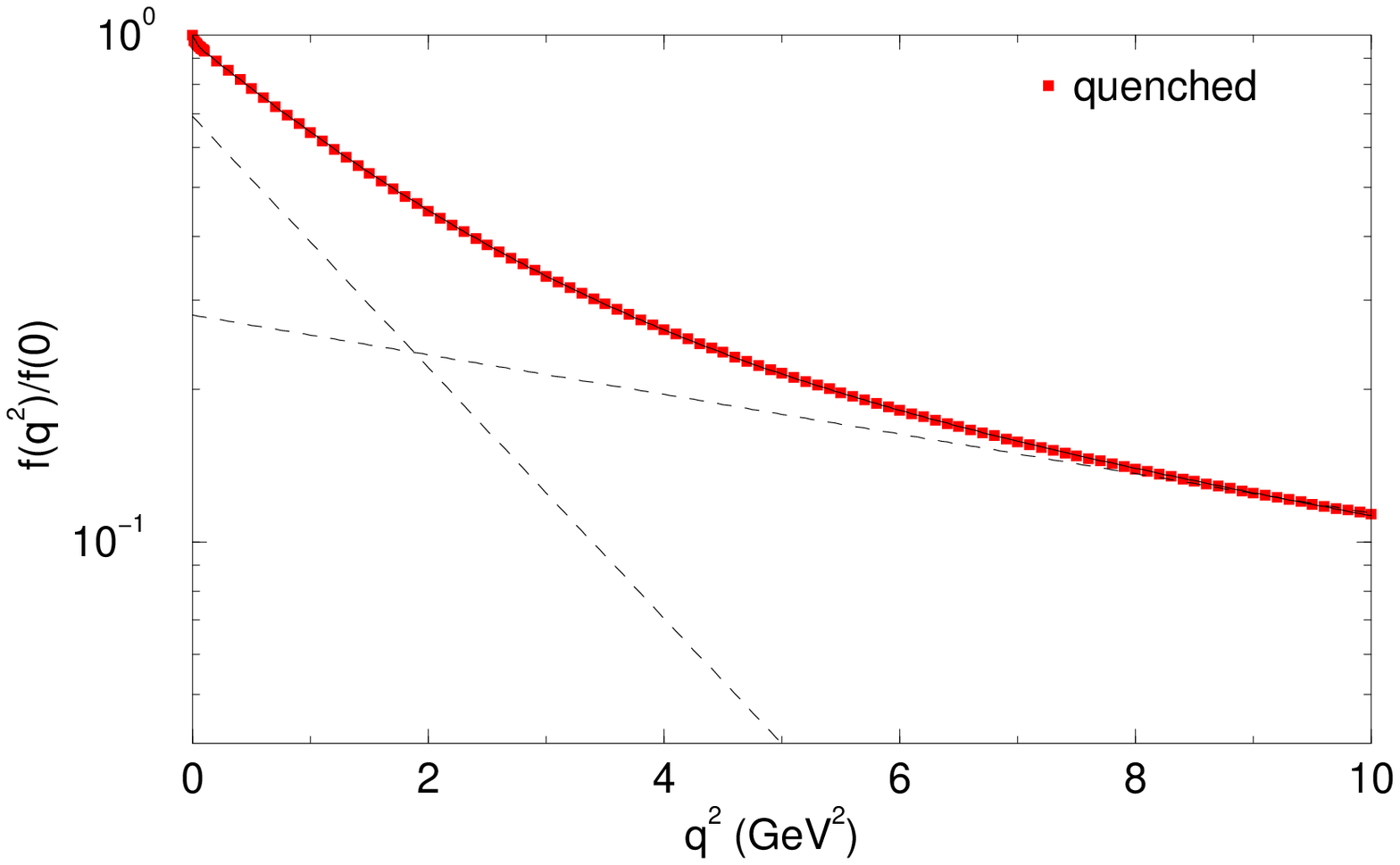}
\end{center}
\begin{center}
\includegraphics[width=8.0cm,height=7.0cm]{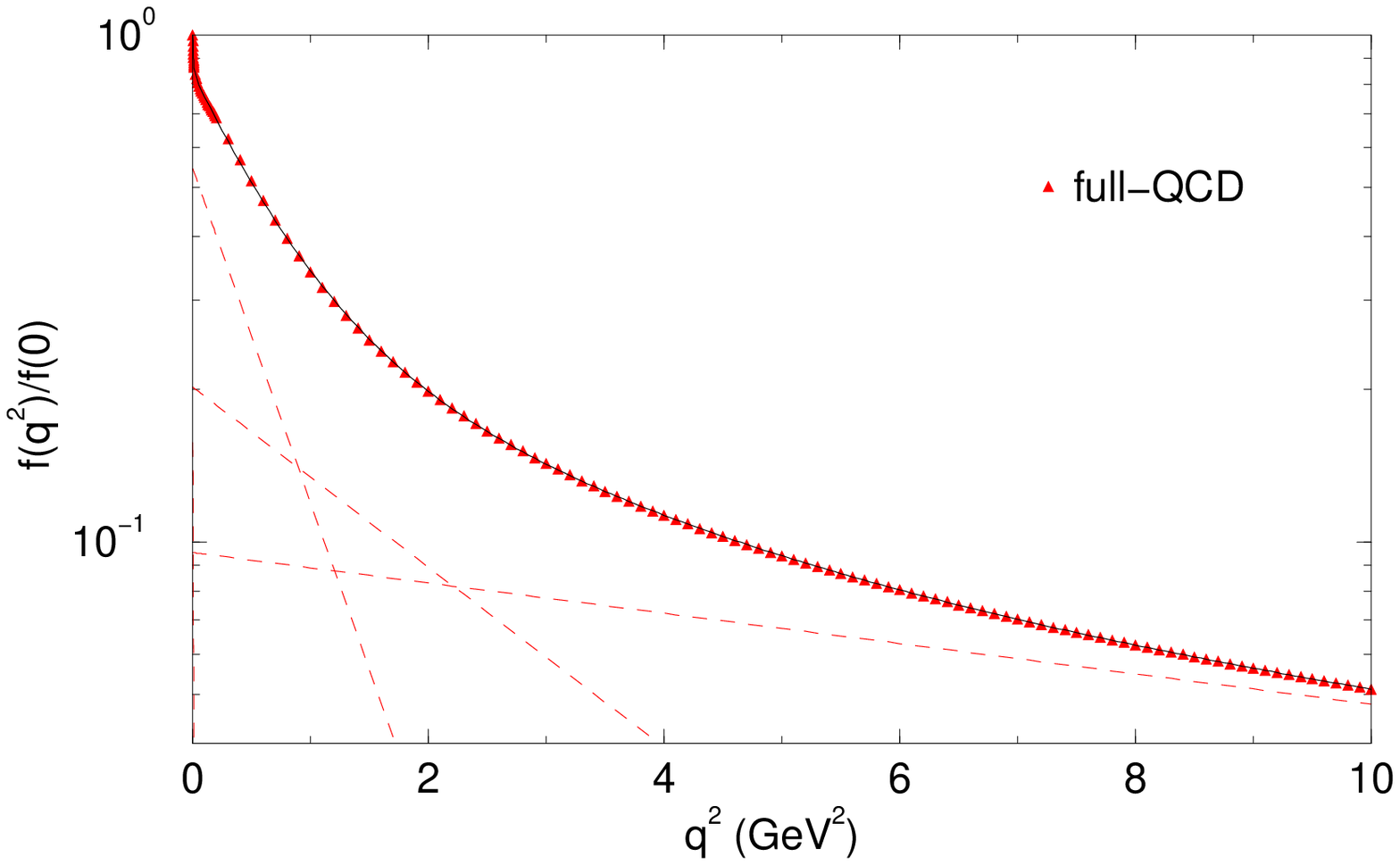}
\caption{Elementary amplitudes from quenched and full QCD and the
exponential components through parametrization (48) \cite{mm03}.}
\end{center}
\end{figure}

Our main conclusions are the following \cite{mm03}: 
(1) the amplitudes decrease smoothly as the
momentum transfer increases and they do not present zeros; (2) the decreasing 
is faster when going from quenched approximation to full QCD (with 
decreasing quark masses), and this effect is associated with the increase 
of the correlation lengths;
(3) the dynamical fermions generate two contributions in the
region of small momentum transfer, which are of the same order at
$q^2 \sim$ 1 GeV$^2$ (only one contribution is present in the
case of quenched approximation).

We understand that result (3) may suggest some kind  of change in
the dynamics at the elementary level, near $q^2 \sim$ 1 GeV$^2$ and at 
asymptotic energies. If that is true, some signal could be expected
at the hadronic level.
One possibility is that this effect can be associated with the position 
of the
dip (or the beginning of the ``shoulder") in the hadronic (elastic) differential
cross section data. The asymptotic condition embodied in our
result indicates that $q^2 \sim$ 1 GeV$^2$ seems to be in agreement with 
the limit of the shrinkage of the diffraction peak, empirically verified when
the energy increases in the region 23 GeV $ \leq \sqrt s \leq $ 1.8 TeV.

\subsubsection{ Correlators from the Instanton Approach}

By means of the stochastic vacuum formalism we also presently 
investigate the elementary amplitudes using 
correlators determined in the context of the \textit{constraint} instanton
approach, developed by A. Dorokhov and collaborators \cite{instanton}.
The basic picture is that of an instanton field dominating
at small distances and decreasing exponentially
at large distances in the physical vacuum.

In Ref. \cite{doro}, we make use of suitable parametrizations for the
correlators and investigate the effects of the contributions
from the short and long range correlations 
in the determination of the full correlator. 
Denoting those contributions 
as $D_I(z)$ and $D_L(z)$, respectively, we introduce a 
dimensionless
parameter $\alpha \equiv \eta_g \rho_c$ in terms of the {\it driven parameter}
$\eta_g$ and the {\it size parameter} $\rho_c$ \cite{doro}.
Since
$\eta_g$ is correlated with the relative contribution of each kind
of correlator, we  consider two extreme cases: 1) equal contributions
(weights 0.5 and 0.5), corresponding to $\alpha = 1.0$; 2) almost pure
instanton contribution (weights 0.99 and 0.01), corresponding to
$\alpha = 0.1$.

In the lack of information on the long range component, and for our purpose,
we consider  parametrization in a Gaussian form \cite{doro}

\begin{eqnarray}
    D_L(z) = \exp \{-(2/2.5)^2 z^2\}. \nonumber
\label{independentlrc}
\end{eqnarray}

For the short range case, $\alpha = 0.1$, we introduce the parametrization,

\begin{eqnarray}
D_I(z)&=&  0.7119\exp\{-(2.403|z|)^2\} \nonumber \\
&+&0.2899\exp\{-(1.485|z|)^2\} \nonumber \\ 
&-& 9.456\times 10^{-3}\exp\{-1.277|z|\} . \nonumber
\label{ddzalfa0pt1}
\end{eqnarray}
The 
full correlator is then determined by

\begin{eqnarray}
    D(z) = 0.99 D_I(z) + 0.01 D_L(z).
\label{src}
\end{eqnarray}

For the long range case, $\alpha = 1.0$, the parametrization takes the form

\begin{eqnarray}
D_I(z)&=& 0.80084\exp\{-(2.3025|z|)^2\} \nonumber \\
&-&3.3846\times 10^{-2}\exp\{-(0.97119|z|)^2\} \nonumber \\ 
&+&0.24225\exp\{-2.7706|z|\}, \nonumber
\label{lrc}
\end{eqnarray}
and for the full correlator we have

\begin{eqnarray}
    D(z) = 0.5 D_I(z) + 0.5 D_L(z).
\label{ddzwithlrc1pt0}
\end{eqnarray}

With the Eqs. (49) and (50) we can
calculate the elementary amplitude through the steps indicated 
in Sec. VI.A. The results are displayed in Fig. (17).

\begin{figure}
\begin{center}
\includegraphics[width=8.0cm,height=7.0cm]{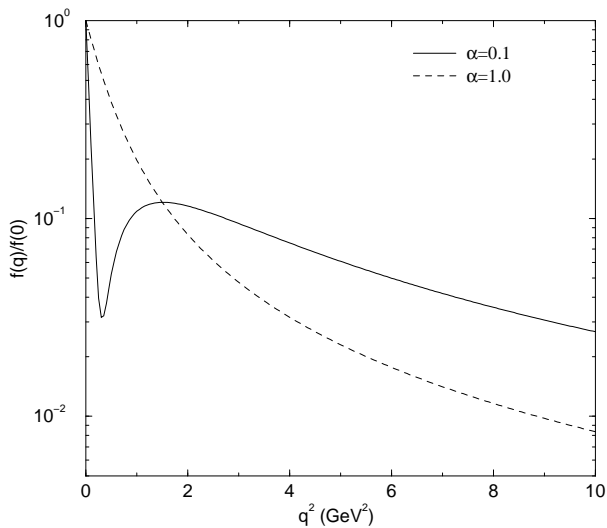}
\caption{Elementary amplitudes from the constrained instanton approach.}
\end{center}
\end{figure}

A central result is that if the contribution from the long
range correlator is small, that is, an almost purely instanton case, the 
corresponding elementary amplitude presents
a minimum. In terms of the associated differential cross section,
this implies a diffractive pattern in the momentum
transfer space, a result already indicated
in some phenomenological approaches \cite{menon,mmt}.
In the case of equal weights the amplitude decreases
monotonically with the momentum transfer.

We conclude that in the context of the instanton approach, the
balance between the contributions of the short and long
range correlators is a crucial point for the determination of the
behavior of the elementary amplitudes. Further investigation
along this line can bring new important insights on the
connection between instanton correlators and the physical
quantities which characterize the high-energy hadronic
scattering.

\section{Perspectives and Outlook}

In this section we outline some perspectives in the area of high-energy
elastic hadron scattering from both experimental and theoretical point of views.
Certainly, some ideas may be biased by our own knowledge and our
own personal view.

\subsection{Experiments}

From the experimental point of view the perspectives
are very optimistic due to 
the new generation of experiments with both accelerators and
cosmic-ray observatories. Let us quickly summarize some projects 
in development.

The upgrade of the Fermilab Tevatron machine, together with upgrades and new
devices in the CDF (Collider Detector Facility) and D0 detectors, are going to
provide improved investigations on $\bar{p}p$ collisions at $\sqrt s \sim 2$
TeV. Although the main purpose of the experiment concerns hard diffraction, it
will be possible to investigate elastic scattering in both high and low $q^2$
regions, the slope parameter, total cross section and single diffraction. 
Of topical importance for soft physics, the new
determination of the total cross section shall possibly bring a solution for
the puzzle represented by the discrepant results around $2$ TeV (E710/E811 and
CDF). 

At the Relativistic Heavy Ion Collider (RHIC) in the Brookhaven National
Laboratory, $pp$ collisions are presently being investigated at energies never
reached before: $\sqrt s : 50\ -\ 500$ GeV. The experiment ``Total and
Differential Cross Sections and Polarization Effects in $pp$ Elastic
Scattering at RHIC'' (pp2pp) plans to investigate both elastic scattering and
diffraction dissociation (single and double), in addition to spin effects.
This will provide the first opportunity for direct comparison
between $pp$ and $\bar{p}p$ scattering at the highest collider energies.

Although in a bit longer term, at the CERN Large Hadron Collider (LHC), the
TOTEM experiment (Total Cross Section, Elastic Scattering and Diffraction
Dissociation at the LHC) is specifically planned to study soft diffractive
physics in $pp$ collisions at $\sqrt s \sim 16$ TeV. In
particular, diffraction dissociation, total cross
section and elastic scattering at large values of the momentum transfer will be
investigated, up to
$q^2 = 10\ -\ 15$ GeV$^2$. That will certainly allow discrimination and
selection of various models and approaches, giving fundamental information at
large momentum transfer; for example, showing the existence or not of
structures. Moreover, this experiment will probably provide a
final answer on the possible differences between $pp$ and $\bar{p}p$ total
cross sections and the correct power $\gamma$ in the $\ln^{\gamma} s$
dependence of $\sigma_{tot}(s)$.

The most energetic event detected in cosmic-ray experiments had $E_{lab} = 3
\times 10^{20}$ eV, corresponding to an energy of $50$ Joules!
Goals of the Auger project are the measurement of arrival direction
and the
energy and mass composition of cosmic rays above $E_{lab} = 10^{19}$ eV.
For $pp$ collisions this means $\sqrt s $ above $140$ TeV, nearly
$10$ times the LHC energy.
In addition to the astrophysical importance of the experiment, the measurement
of the longitudinal development of showers will provide severe tests on
hadronic interaction models. As a consequence, among others, the puzzles
concerning the extraction of $pp$ cross section from $p-air$ production cross
section may receive better insights, allowing more precise determinations and
at energies possibly never to be reached by accelerator machines.

\subsection{Theory}

Elastic hadron scattering (and soft diffractive processes in general)
is a long distance phenomena and therefore we expect and look for a theoretical
treatment via non-perturbative QCD. Despite all the difficulties mentioned along
this manuscript, we understand that two approaches deserve special attention.

One of them concerns the approach by Nachtmann and the Stochastic Vacuum Model
(Sec. VI.A) \cite{otto1,rev}.
Although under restrictive kinematic conditions (momentum transfer of the
order or below 1 GeV$^2$, and asymptotic energies, $s \rightarrow \infty $) the
formalism has provided interesting results in the investigation of
the physical quantities that characterize the elastic $pp$
and $\bar{p}p$ scattering \cite{dfk,berger}, in special  the works
by Ferreira and Pereira, connecting experimental observables and QCD parameters 
\cite{fp}.
Attempts to implement the energy dependence in pure QCD grounds
may be an important task for the near future.

The other approach is associated with evidences for finite gluon
propagator and running coupling in the infrared region and
that is the case in some classes of solutions of the
nonperturbative Schwinger-Dyson equations. In particular,
in the solution proposed by Cornwall \cite{cornwall} the gluon
acquires a dynamical mass leading to a freezing of the coupling
constant in the infrared region. As referred before, Natale and
collaborators \cite{natale} have discussed several phenomenological tests,
reaching
interesting results which have permitted the development of the
formalism and the selection of adequate basic inputs.
That opens a new way to investigate long distance phenomena
with a finite calculational approach.

We also understand that the connections between soft and semi-hard processes, 
typical of QCD-based models in the eikonal context, may bring
new insights for the development of adequate calculational schemes in
the nonperturbative treatment of high-energy elastic collisions.

\section{Summary and Final Remarks}

Despite its simplicity,
elastic hadron scattering constitutes a topical problem in
high-energy Physics. Although the bulk of experimental data 
can be efficiently described in different phenomenological
contexts,
we are still facing the lack of a treatment and of
a reasonable understanding of these processes based exclusively
on QCD.

Our main strategy in investigating elastic scattering has been to look for 
descriptions based 
on the high-energy principles and theorems from  Axiomatic
Quantum Field Theory and, simultaneously, attempting to extract  
``empirical" information from all the experimental
data available. Tests of discrepant data and their influence
on the extracted information play a central role.
In that way we hope to get 
feedbacks for theoretical development in nonperturbative
and semi-hard QCD.
We can summarize our main recent results as follows.

In the context of the \textit{Analytic Approach},
we have investigated the effects of discrepant experimental information
on the total cross sections in both accelerator and cosmic-ray
energy regions. By means of analytical fits, we have obtained extrema bounds
for the soft Pomeron intercept, namely $\alpha^{upper}_{\tt I\!P}(0) = 1.109$
and $\alpha^{lower}_{\tt I\!P}(0) = 1.081$. 
We have also obtained novel constrained bounds for the intercept from spectroscopy
data (fitted Regge trajectories from Chew-Fautschi plots) and extended
the analysis to several reactions.
That information on the Pomeron bounds may be important for
phenomenological developments and projects for new experiments.
We have also shown that the presence of the Odderon in the real
part of the elastic hadronic amplitude is not forbidden by the
bulk of experimental data on $\sigma_{tot}$ and $\rho$. In particular,
the fit with the Kang-Nicolescu parametrization has
indicated a crossing in $\rho(s)$, with $\rho_{pp}$ becoming
greater than $\rho_{\bar{p}p}$ above $\sqrt s$ = 70 GeV. That
parametrization predicts $\rho_{pp}(\sqrt s = 200 \mathrm{GeV})
= 0.134 \pm 0.005$ (RHIC regions). Detailed investigation
on the applicability of DDR have shown that, once the subtraction
constant is used as a free fit parameter, the DDR is
equivalent to the IDR with finite lower limit ($s_0 = 2m^2$).
That result was obtained for the class of entire functions in the
logarithm of the energy (typical of analytic models).

In the context of \textit{Model Independent Analyses}, we have investigated the 
correlations
between the experimental data on total cross section and the slope parameter.
The parametrization introduced is based on the empirical behavior of these
quantities and extrapolations to cosmic-ray energies may be useful
in the determination of proton-proton total cross sections
from proton-air cross sections.
By means of a novel model independent fit procedure to the differential cross 
section data,
we have found statistical evidence for eikonal zero in the momentum
transfer space and that the position of the zeros decreases as the energy 
increases.
The zero position shows agreement with the result recently obtained for
the electromagnetic form factor and inferred from elastic electromagnetic
$e^{-}p$ scattering (polarization transfer experiments). Since our analysis 
concerns only hadronic interactions
the results may bring new insights in the investigation of electromagnetic
and hadronic form factors. We are also treating analytic fits with
free parameters depending on the energy so as to develop a model
independent predictive approach.

In the context of \textit{Eikonal Models} we have obtained connections between
the elastic and inelastic channel by means of the geometrical picture,
and correlating $pp$ and $\bar{p}p$ scattering with contact interactions
simulated by $e^{-}e^{+}$ distributions and multiplicities. With the class
of QCD-based or QCD-inspired models we have developed novel gluonic
contributions by means of two approaches. One is based in solutions
of the Schwinger-Dyson equations characterized by the dynamical gluon mass. 
The other one is intended to take into account the $Q^2$- scale in updated gluonic 
distribution functions. Certainly, the two approaches are not independent, and 
we presently investigate their simultaneous implementation.

In the context of the \textit{Stochastic Vacuum Model}, we have obtained
novel results for the elementary scattering amplitudes, making use of
correlators determined either from lattice QCD or from constrained instantons.
In both cases the elementary amplitudes present no zeros (change of
sign in the momentum transfer space). In the context of eikonal models,
in which the eikonal is expressed in terms of the elementary amplitudes
and form factors, Eq. (38), this result corroborates the interpretation of 
the zero in the hadronic form factor, and therefore, its dependence on the energy.

\begin{acknowledgments}

I am thankfull to Y. Hama and F.S. Navarra for the invitation
to present this review. For fruitful comments and discussions along these
Workshops I am particularly grateful to E. Ferreira, Y. Hama and T. Kodama.
I am deeply thankful to those who have developed research programs
under my supervision and are co-authors of all the works 
reviewed here: R.F. \'Avila,
P.C. Beggio, S.D. Campos, P.A.S. Carvalho, E.G.S. Luna, A.F. Martini, 
J. Montanha, and D.S. Thober.
I am also thankful to A. Di Giacomo, P. Valin, A. Dorokhov, A.A. Natale 
and R.C. Rigitano for valuable discussions.
This work was supported by FAPESP (Contract N. 00/04422-7).

\end{acknowledgments}


\end{document}